\pdfoutput=1
\newcommand*{\ATLASLATEXPATH}{}
\documentclass[cernpreprint,USenglish,texlive=2011,txfonts]{\ATLASLATEXPATH atlasdoc}

\usepackage{\ATLASLATEXPATH atlaspackage}
\usepackage{\ATLASLATEXPATH atlasbiblatex}

\usepackage{\ATLASLATEXPATH atlascontribute}

\usepackage{\ATLASLATEXPATH atlasphysics}

\usepackage{multirow}
\usepackage{rotating}
\usepackage{tabularx}
\usepackage{epstopdf}

\usepackage{dileptonAsymmetry-defs}

\AtlasTitle{Measurements of the charge asymmetry in top-quark pair production in the dilepton final state at $\sqrt{s}=8$~\TeV~with the ATLAS detector}

\author{The ATLAS Collaboration}

\AtlasRefCode{TOPQ-2015-08}

\PreprintIdNumber{CERN-EP-2016-051}

\AtlasJournal{Phys.\ Rev.\ D.}

\AtlasJournalRef{Phys. Rev. D 94, 032006 (2016)}
\AtlasDOI{10.1103/PhysRevD.94.032006}

\AtlasAbstract{
Measurements of the top--antitop quark pair production charge asymmetry
in the dilepton channel, characterized by two high-${p}_{\rm{T}}$ leptons (electrons or
muons), are presented using data
corresponding to an integrated luminosity of \lumi from $pp$
collisions at a center-of-mass energy $\sqrt{s} = 8$~\TeV~collected
with the ATLAS detector at the Large Hadron Collider
at CERN. Inclusive and differential measurements as a function of
the invariant mass, transverse momentum, and longitudinal boost of
the $t\bar{t}$  system
are performed both in the full phase space and in a fiducial
phase space closely matching the detector acceptance.
Two observables are studied: \Acll\ based on the selected leptons and \Ac based on the
reconstructed \ttbar\ final state. The
inclusive asymmetries are measured in the full phase space to be
$\Acll = \Acllcomb$ and
$\Ac  =  \Accomb$,
 which are in agreement with the Standard Model predictions of $\Acll =
\Aclltheory$ and $\Ac = \Actttheory$.
}

\hypersetup{pdftitle={ATLAS document},pdfauthor={The ATLAS Collaboration}}

\begin{document}

\maketitle

\tableofcontents

\section{Introduction}
\label{sec:intro}
The top quark is the heaviest known elementary particle. Its large mass suggests that it may play a
special role in theories of physics beyond the Standard Model (BSM)~\cite{Hill:1994hp,Lillie:2007yh,AXI,ZP,KK}. Such a role could be elucidated via precision tests of the Standard Model (SM) in large data samples of top--antitop quark pair (\ttbar) events collected at the Large Hadron Collider (LHC) in
proton--proton ($pp$) collisions. One such test is the measurement of the charge asymmetry. The production of \ttbar~pairs at hadron colliders is symmetric under charge
conjugation at leading order (LO) in quantum chromodynamics (QCD),
i.e., the probability of a top quark flying in a given direction is the
same as for an antitop quark~\cite{Aguilar-Saavedra:2014kpa}.  At
next-to-leading order (NLO) in QCD, an asymmetry arises from
interference between different Feynman diagrams~\cite{AXI}. In particular, interference
between the Born and one-loop diagram of the $q\bar{q} \to t\bar{t}$~processes
and between $q\bar{q} \to t\bar{t}g$~diagrams with initial-state and
final-state radiation (ISR and FSR) processes lead to a charge asymmetry. In the
\ttbar~rest frame, this
asymmetry causes the top quark to be preferentially emitted in the
direction of the initial quark, and causes the antitop quark to be
emitted in the direction of the initial antiquark. The size of the
asymmetry can be enhanced by
contributions beyond the SM, for example, \ttbar~production via the
exchange of new heavy particles such as axigluons~\cite{AXI}, heavy
$Z$ particles~\cite{ZP}, or colored Kaluza--Klein excitations of the
gluon~\cite{KK}.

Inclusive and differential measurements of the \ttbar~asymmetry were
first performed at the Tevatron proton--antiproton collider,
where forward-backward asymmetries were measured. Several
measurements were reported by the CDF
and D0 experiments~\cite{CDF2,CDF1,Aaltonen:2012it,D02,D03,Abazov:2015fna}
in dileptonic and semileptonic \ttbar~events. For these
measurements, the direction of the initial quark can be assumed
to be the direction of the proton, and the direction of the
antiquark that of the antiproton, which yields straightforward
access to the asymmetry. Initial tension between these measurements
and theory predictions have
been reduced with the latest SM calculations at next-to-next-to-leading order
(NNLO) QCD~\cite{Czakon:2014xsa}.

Since the start of the LHC, measurements of $t\bar{t}$ charge asymmetries have
been performed by the ATLAS and CMS experiments. Two features
complicate the measurement of the asymmetry
at the LHC: in proton--proton collisions the initial state is symmetric, so
there is no \ttbar~forward-backward asymmetry,
and the dominant production mechanism is
gluon fusion, which is symmetric under charge conjugation to all
orders in perturbative QCD. However, valence quarks carry on average a
larger fraction of the proton momentum than sea antiquarks, hence top  antiquarks
 produced through quark--antiquark annihilation are more central than top quarks~\cite{Kuhn:1998jr}.
By using differences between the absolute
rapidity of the top and antitop quarks, ATLAS and CMS performed
measurements of the charge asymmetry in dileptonic and semileptonic
events at $\sqrt{s} = 7$~\TeV~and $8$~\TeV~\cite{atlas7tev,atlas7tevljets,cms7tevljets,cms7tevdil,atlas8tevljets,cms8tevljets,Khachatryan:2015mna,Khachatryan:2016ysn}.
All asymmetry measurements at the LHC show good agreement with the SM
prediction~\cite{Bernreuther:2012sx}, which is approximately an order of magnitude smaller than the predicted
asymmetry at the Tevatron.

In this article, new measurements of the charge asymmetry are presented
using dileptonic \ttbar~events at $\sqrt{s} = 8$~\TeV. The dileptonic channel is characterized by two charged leptons (denoted $\ell=e,\mu$), coming from either a direct
vector boson decay or through an intermediate $\tau$ lepton decay. Two different observables are used, based either on the the selected leptons or the reconstructed  $t\bar{t}$ final state. Inclusive and
differential measurements as a function of the invariant mass of
the \ttbar~system ($m_{t\bar{t}}$), the transverse momentum of
the \ttbar~system ($p_{\rm{T},t\bar{t}}$), and the absolute value of the boost of the
\ttbar~system along the beam axis ($\beta_{z,t\bar{t}}$) are performed.
The inclusive and differential measurements are performed in the full phase space as well as in a
fiducial volume based on the detector acceptance and selection
requirements, using particle-level objects.
The measurement in the fiducial region does not rely on extrapolating
 to regions of phase space that are not within the detector acceptance, while the
full phase space measurement has the benefit of being comparable to
theoretical calculations at the parton level, including BSM models.

In Sec.~\ref{sec:detector}, a brief description of the ATLAS detector is
given. Section~\ref{sec:montecarlo} describes the data and Monte Carlo (MC) samples,
 and Sec.~\ref{sec:selection} the event selection and background estimation.
The observables are described in
Sec.~\ref{sec:observables}. Section~\ref{sec:analysis} outlines the
measurement methods, including a description of
the \ttbar~reconstruction, the definition of fiducial volume, and a
description of the unfolding procedure. In
Sec.~\ref{sec:systematics}, the sources of systematic uncertainties
affecting the measurements are discussed, and results are provided in
Sec.~\ref{sec:result}. Finally, conclusions are given in Sec.~\ref{sec:conclusions}.
\section{ATLAS detector}
\label{sec:detector}
The ATLAS detector~\cite{ATLAS} at the LHC covers nearly the entire solid angle around the interaction
point.\footnote{ATLAS uses a right-handed coordinate system with its origin at the nominal interaction point (IP) in
the center of the detector and the $z$-axis along the beam pipe. The $x$-axis points from the IP to the center of the LHC ring, and the $y$-axis points upward. Cylindrical coordinates $(r, \phi)$ are used in the transverse
plane, $\phi$ being the azimuthal angle around the beam pipe.}
It consists of an inner tracking detector surrounded by a thin superconducting solenoid, electromagnetic and hadronic calorimeters, and a muon spectrometer incorporating
 superconducting toroid magnets.

The inner-detector system is immersed in
a $2$ T axial magnetic field and provides charged-particle tracking in
the pseudorapidity\footnote{The pseudorapidity is defined in terms of
the polar angle $\theta$ as $\eta = - \ln[\tan \theta/2]$, while the
rapidity $y$ is defined as $y = - (1/2) \ln[(E + p_z)/(E - p_z)]$.} range $|\eta| <2.5 $. A high-granularity silicon pixel detector covers the interaction region and
provides typically three measurements per track. It is surrounded by a silicon microstrip tracker
designed to provide four two-dimensional measurement points per track. These silicon detectors
are complemented by a transition radiation tracker, which enables radially extended
track reconstruction up to $|\eta| =2.0 $. The transition radiation tracker also provides electron
identification information based on the fraction of hits (typically 30 in total) exceeding an
energy-deposit threshold corresponding to transition radiation.

The calorimeter system covers the pseudorapidity range $|\eta| <4.9 $. Within the region
$|\eta| <3.2 $, electromagnetic calorimetry is provided by barrel and endcap high-granularity
lead/liquid-argon (LAr) electromagnetic calorimeters, with an additional thin LAr presampler
covering $|\eta| <1.8 $ to correct for energy loss in the material upstream of the calorimeters.
Hadronic calorimetry is provided by a steel/scintillator-tile calorimeter, segmented into
three barrel structures within $|\eta| <1.7 $, and two copper/LAr hadronic endcap calorimeters.
The solid angle coverage is completed with forward copper/LAr and tungsten/LAr
calorimeters used for electromagnetic and hadronic measurements, respectively.

The muon spectrometer comprises separate trigger and high-precision tracking chambers
measuring the deflection of muons in a magnetic field generated by superconducting
air-core toroids. The precision chamber system covers the region $|\eta| <2.7 $ with drift tube chambers, complemented by cathode strip chambers. The muon trigger system covers the range of $|\eta| <1.05$ with resistive plate chambers
in the barrel, and the range of $1.05< |\eta| <2.4$ with thin gap chambers in the endcap regions.

A three-level trigger system is used to select interesting events. The Level-1 trigger is
implemented in hardware and uses a subset of detector information to reduce the event rate
to a design value of at most 75 kHz. This is followed by two software-based trigger levels,
which together reduce the event rate to about 300 Hz.

\section{Data and Monte Carlo samples}
\label{sec:montecarlo}
The data used for this analysis were collected during the 2012 LHC
running period at a center-of-mass energy of $8$ \TeV. After
applying data-quality selection criteria, the data sample used in the
analysis corresponds to an integrated luminosity of $20.3$
$\rm{fb}^{-1}$.

For the modeling of the signal processes and most background
contributions, several MC event generators are used. The main
background contribution in this measurement comes from Drell--Yan
production of $Z/\gamma^{*} \rightarrow \ell \ell$, which is estimated
by a combination of simulated samples modified with corrections
derived from data, as described in
Sec.~\ref{sec:selection}. The smaller
contributions from diboson ($WW$, $ZZ$, and $WZ$) and single-top-quark ($Wt$
channel) production are evaluated purely via MC simulations. Further background contributions can arise from events including a jet or a
lepton from a semileptonic hadron decay misidentified as an isolated
charged lepton as well as leptons from photon conversions, together
referred to as ``fake leptons''.
 This contribution is estimated using simulated
 samples, modified with corrections derived from data. The samples mentioned above together with simulated samples of $t\bar{t}+W/Z$, $t$-channel of single-top-quark production,
 $W$+jets, and $W$+$\gamma$+jets are included in the estimation. The estimation
 procedure is described in Sec.~\ref{sec:selection}.

The nominal $t\bar{t}$ signal sample is generated at NLO in QCD using \powheg
(version 1, r2330)~\cite{Alioli:2010xd,Frixione:2007vw,Nason:2004rx}
and the CT10~\cite{ct10} parton distribution function (PDF) set, setting the
$h_{\rm{damp}}$ parameter to the top-quark mass of $172.5 \gev$. The
$h_{\rm{damp}}$
parameter is the resummation scale that is used in the damping
function, which is designed to limit the resummation of higher-order
effects at large transverse momentum without spoiling the NLO accuracy
of the cross section.
The parton shower, hadronization, and underlying event are simulated using \pythia (version 6.427)~\cite{Sjostrand:2006za}
with the CTEQ6L1 PDF~\cite{Pumplin:2002vw} and the corresponding set of
tunable parameters (Perugia 2011C tune~\cite{Skands:2010ak})
intended to be used with this PDF.
The $t\bar{t}$ cross section for $pp$ collisions at a center-of-mass energy of $8$~\TeV~ is set to
$\sigma_{t\bar{t}}= 253^{+13}_{-15}$~pb, calculated at NNLO in QCD including resummation of next-to-next-to-leading logarithmic (NNLL) soft gluon terms with top++2.0~\cite{Cacciari:2011hy,Beneke:2011mq,Baernreuther:2012ws,Czakon:2012zr,Czakon:2012pz,Czakon:2013goa,Czakon:2011xx}. The PDF and $\alpha_{\rm{S}}$ uncertainties were calculated using the PDF4LHC prescription~\cite{Botje:2011sn} with the MSTW2008 $68\%$ CL NNLO~\cite{Martin:2009iq,Martin:2009bu}, CT10 NNLO~\cite{Lai:2010vv,Gao:2013xoa}, and NNPDF2.3 5f FFN~\cite{Ball:2012cx} PDF sets, and added in quadrature to the scale uncertainty.  

Single-top-quark production in the $Wt$ channel is simulated using \powheg
with \pythia (version 6.426) and the CT10 (NLO) PDF set. The cross
section of $ 22.3 \pm 1.5$~pb is estimated at approximate NNLO in
QCD including resummation of NNLL terms~\cite{Kidonakis:2010ux}.
The parton shower, hadronization, and underlying event are simulated
by \pythia  using the Perugia 2011C tune. The Drell--Yan process is modeled using \alpgen (version 2.14)~\cite{Mangano:2002ea} interfaced
with \pythia with the CTEQ6L1~\cite{Pumplin:2002vw} PDF set using
the MLM matching scheme.  Its
heavy-flavor component is included in the matrix element calculations
to model the $Z/\gamma^{*} + b\bar{b}$ and $Z/\gamma^{*} + c\bar{c}$
processes. Diboson processes ($WW$, $ZZ$, and $WZ$) are simulated using \alpgen interfaced with \herwigjimmy (version
4.31)~\cite{Corcella:2000bw,JIMMY} with the
CTEQ6L1~\cite{Pumplin:2002vw} PDF set for parton fragmentation~\cite{Mangano:2001xp}.
The only exceptions are the same-charge $W^{+(-)}W^{+(-)}$ samples, which
are simulated using \madgraph (version 5.1.4.8)~\cite{Alwall:2011uj}
interfaced with \pyEight (version 8.165)~\cite{Sjostrand:2007gs}.  The
samples are normalized to the reference  NLO QCD prediction, obtained
using the MCFM generator~\cite{Campbell:1999ah}.
The associated production of a $t\bar{t}$ pair with a vector boson ($t\bar{t}Z$ and $t\bar{t}W$) is simulated with \madgraph interfaced with \pyEight and normalized to NLO cross-section calculations~\cite{Campbell:2012dh,Garzelli:2012bn}. The $W$+jets events are simulated
using \alpgen interfaced with \pythia  and the $W$+$\gamma$+jets process
is simulated using \alpgen interfaced with \jimmy.

To model the LHC environment properly, additional inelastic $pp$
collisions are generated with \pyEight\ and overlaid on the hard
process. All the simulated samples are then processed through a simulation of the ATLAS detector~\cite{Aad:2010ah}. For most of the samples, a full simulation based
on GEANT4~\cite{geant4} is used. Some of the samples used to evaluate the generator modeling uncertainties are obtained using a faster detector simulation where only the calorimeter simulation is modified and relies on parametrized showers~\cite{ATL-PHYS-PUB-2010-013}. The simulated events are passed through
the same reconstruction and analysis chain as data.

\section{Event selection and background estimation}
\label{sec:selection}
In order to enrich the data sample in dileptonic \ttbar~events,
requirements are imposed on reconstructed charged leptons (electrons and
muons), jets, and the missing transverse momentum. Three different final
states are considered in the analysis: events with two electrons in
the final state ($ee$), with one electron and one muon ($e\mu$), and
 with two muons ($\mu\mu$).

Electron candidates are reconstructed from an electromagnetic
calorimeter energy deposit matched to a track in the inner
detector and must pass the  likelihood-based ``medium'' identification
requirements~\cite{Aad:2014fxa}. They are required to have
transverse momentum $p_T > 25$ \GeV~and must also lie in the region
\mbox{$|\eta_{\rm cl}| < 2.47$}, where $\eta_{\rm cl}$ is the pseudorapidity
of the calorimeter energy cluster associated with the electron,
excluding the transition region between the calorimeter
barrel and endcaps $1.37 < |\eta_{\rm cl}| <
1.52$. Moreover, electrons are required to be isolated from
surrounding activity in the inner detector. The
scalar sum of the track $p_{\rm{T}}$ within a cone of
$\Delta R = \sqrt{(\Delta \eta)^2+(\Delta \phi)^2} = 0.3$ (excluding the track of
the electron itself) divided by the electron $p_{\rm{T}}$ should be less than 0.12.

Muon candidates are reconstructed using combined information from the muon
spectrometer and the inner detector~\cite{Aad:2014rra}. They are required to
have $\pt > 25 \GeV$ and $|\eta| < 2.5$. In addition, muons are
required to satisfy track-based \pt-dependent isolation
criteria. The scalar sum of the track \pt\ within a cone of size \mbox{$\Delta R =
10~\textrm{\GeV}/p_{\rm{T}}^{\mu}$} around the muon (excluding the muon track
itself) must be less than 5\% of the muon \pt\ ($p_{\rm{T}}^{\mu}$).
Both the electrons and muons have to be consistent with the primary vertex,\footnote{The primary vertex is defined as the
 reconstructed vertex with at least five associated tracks (of $p_T >$
 0.4 \GeV) and the highest sum of the squared transverse momenta
of the associated tracks.} by requiring the absolute value of the
longitudinal impact parameter to be less than 2~mm.

Jets are reconstructed from clustered energy deposits in the
electromagnetic and hadronic calorimeters, using the anti-$k_{t}$~\cite{Cacciari:2008gp} algorithm with a radius
parameter $R = 0.4$.
The measured energy of the jets is corrected to the hadronic scale using \pt-
and $\eta$-dependent scale factors derived from simulation and validated in
data~\cite{Aad:2014bia}.
After the energy correction, the jets are required to have
$\pt > 25 \GeV$, to be in the pseudorapidity range $|\eta| < 2.5$, and to have a jet vertex fraction
$|\textrm{JVF}| > 0.5$ \cite{Aad:2015ina} if $\pt < 50 \GeV$.
The jet vertex fraction is defined as the summed scalar $p_{\rm T}$ of the tracks associated with both the jet and the primary vertex divided by the summed scalar $p_{\rm T}$ of all tracks in the jet.
The jet that is the closest to a selected electron is removed from
the event if their separation is $\Delta R < 0.2$.
After this jet overlap removal, electrons and muons that are within a cone of $\Delta R = 0.4$ around the closest jet are removed.
Jets containing $b$-hadrons are identified ($b$-tagged) using a
multivariate algorithm
(MV1)~\cite{Aad:2015ydr}. This is a neural-network-based algorithm that makes use of track impact parameters
and reconstructed secondary vertices. Jets are identified as $b$-tagged
jets by requiring the MV1 output discriminant to be above a certain
threshold value. This value is chosen such that the overall tagging
efficiency for $b$-jets with $\pt > 20 \GeV$
and $|\eta| < 2.5$ originating from top-quark decays in dileptonic MC
$t\bar{t}$ events is 70\%.
The rejection factor for jets originating from gluons and light quarks
is about 130, while for $c$-quarks it is about 5.

The magnitude of the missing transverse momentum (\met) is calculated from the
negative vector sum of all calorimeter energy deposits and the momenta of
muons~\cite{Aad:2012re}.
The calculation is refined by the application of the object-level
corrections for the contributions arising from identified electrons and
muons.

Events recorded with single-lepton triggers
($e$ or $\mu$) under stable beam conditions with all detector
subsystems operational are considered.  The transverse momentum thresholds are $24$ \GeV~for isolated single-lepton triggers and $60$ ($36$) \GeV~for nonisolated single-electron (single-muon) triggers. The nonisolated triggers are
used to select events that fail the isolation requirement at trigger
level but pass it in the offline analysis.
In all three final states, exactly two isolated leptons with opposite
charge and an invariant mass $m_{\ell  \ell} > 15$ \GeV~are required,
together with at least two jets.
In the same-flavor channels ($ee$ and $\mu\mu$),
the invariant mass of the two charged leptons is required to be
outside of the $Z$ boson mass window such
that \mbox{$|m_{\ell \ell}-m_Z|>10$ \GeV}.
Furthermore, it is required that $\met > 30$ \GeV~and at
least one of the jets must be $b$-tagged.
These requirements suppress the dominant background contribution from
Drell--Yan production of $Z/\gamma^{*} \rightarrow \ell \ell$ and also
suppress diboson backgrounds.
In the $e\mu$ channel, the background contamination is much smaller
and the background suppression is achieved by
requiring the scalar sum of the $p_{\rm{T}}$ of the two leading jets and
leptons $(H_{\rm{T}})$ to be larger than $130$ \GeV.
The event selection requirements are summarized in Table~\ref{tab:cuts}.

\begin{table}[h]
  \centering
  \caption{The summary of the event selection requirements applied in different channels.\label{tab:cuts}}
  \begin{tabular}{lcc}
    \hline
    \hline
    Requirements      & $ee/\mu\mu$  & $e\mu$  \\ \hline
Leptons            &  2 & 2\\
Jets               &  $\ge 2$ & $\ge 2$\\
$m_{\ell  \ell}$      &  $> 15 $ \GeV & $> 15 $ \GeV \\
$|m_{\ell \ell}-m_Z|$ & $> 10 $ \GeV & -- \\
$\met$             & $> 30 $ \GeV  & --  \\
$b$-tagged jets    & $\ge 1$       & --  \\
$H_{\rm{T}}$              & --            & $> 130 $ \GeV  \\
    \hline
    \hline
  \end{tabular}
\end{table}


The modeling of Drell--Yan events in the
same-flavor channels with $\met > 30$ \GeV~may not be
accurate in simulation due to the mismodeling of the $\met$ distribution.
Moreover, after applying the $b$-tagging requirement, a large
contribution to the background comes from the associated production of
$Z$ bosons with heavy-flavor jets, which is not well predicted by MC
simulation.  The first source of mismodeling depends on the reconstructed objects and is therefore different in each channel. The second source is a limitation of the MC simulation and is expected to be the same in both channels. Thus, the normalization of the
inclusive and heavy-flavor component of the Drell-Yan background in the
same-flavor channels is computed simultaneously using
data in two control regions with three scale factors.
Two scale factors are applied to all Drell--Yan events to take into
account the mismodeling from the $\met$ requirement (one in the $ee$
and one in the $\mu\mu$ channel) while another is applied only to
$Z$+heavy-flavor events.
The control regions are defined using the standard selection described
previously but inverting the $m_{\ell \ell}$ cut to be within the $Z$
mass window.
The first control region is defined without the $b$-tagging requirement
while the second is defined with at least one $b$-tagged jet. The simulated $m_{\ell \ell}$ distribution in these control regions is simultaneously fit to the data and the scale factors are extracted.
The scale factors derived in these two regions are $0.927 \pm 0.005$ and $0.890 \pm 0.004$ for the $ee$ and $\mu\mu$ channels, respectively, and $1.70\pm0.03$ for the heavy-flavor component. The $Z \rightarrow \tau\tau$ process in the $e\mu$ channel is estimated using MC simulation only: no data-driven correction is applied since neither the $\met$ requirement nor $b$-tagging requirement are applied to this channel.

The background arising from misidentified and nonprompt (NP) leptons is
determined using both MC simulation and data.
The dominant sources of these fake leptons are semileptonic $b$-hadron decays, long-lived
weakly decaying states (such as $\pi^{\pm}$ or $K^{\pm}$ mesons), $\pi^{0}$
showers, photons reconstructed as electrons, and electrons from photon conversions.
$W$+jets, $W$+$\gamma$+jets, \ttbar, $t\bar{t}Z$, $t\bar{t}W$, Drell--Yan, single-top-quark, and diboson production
are taken into account for the estimation of this background.
Multijet events do not contribute significantly to this background,
since the probability of having two jets misidentified as isolated leptons is very small.
The shapes of the kinematic distributions are taken from simulated events where
at least one of two selected leptons is required not to be matched
with the MC generator-level leptons.
Scale factors are derived from data in order to adjust the normalization.
A control region, enriched in fake leptons, is defined by applying the same cuts as
for the final selection but requiring the two leptons to have the same charge.
The shapes of the distributions for various kinematic variables of
leptons, jets, and \met are checked and found to be well modeled
in the MC simulation.
The scale factors are derived in this region by comparing data and
simulation and are then applied to the simulated events in the signal region.
The scale factor is $1.2 \pm 0.3$ in the $ee$ channel, $1.1 \pm 0.2$ in the  $e\mu$
channel, and $3.7 \pm 0.8$ in the $\mu\mu$ channel, where the
uncertainties are statistical. The sources of misidentified muons, such as heavy-flavor decays, are quite different from those of misidentified electrons. The large difference between the scale factor for the $\mu\mu$
 and the $e\mu$ channel is mainly due to the $b$-tagging requirement, that is
 applied only in the $\mu\mu$ channel. However, the shapes of the distributions of the relevant kinematic variables
 in the $\mu\mu$ channel are cross-checked in control regions and found
 to be consistent with the distributions from a purely data-driven
 method.
The systematic uncertainties of both Drell--Yan background and the
background due to events from misidentified and nonprompt leptons are
discussed in detail in Sec.~\ref{sec:syst_background}.

\begin{figure}[!htbp]
 \begin{center}
   \includegraphics[width=0.49\textwidth]{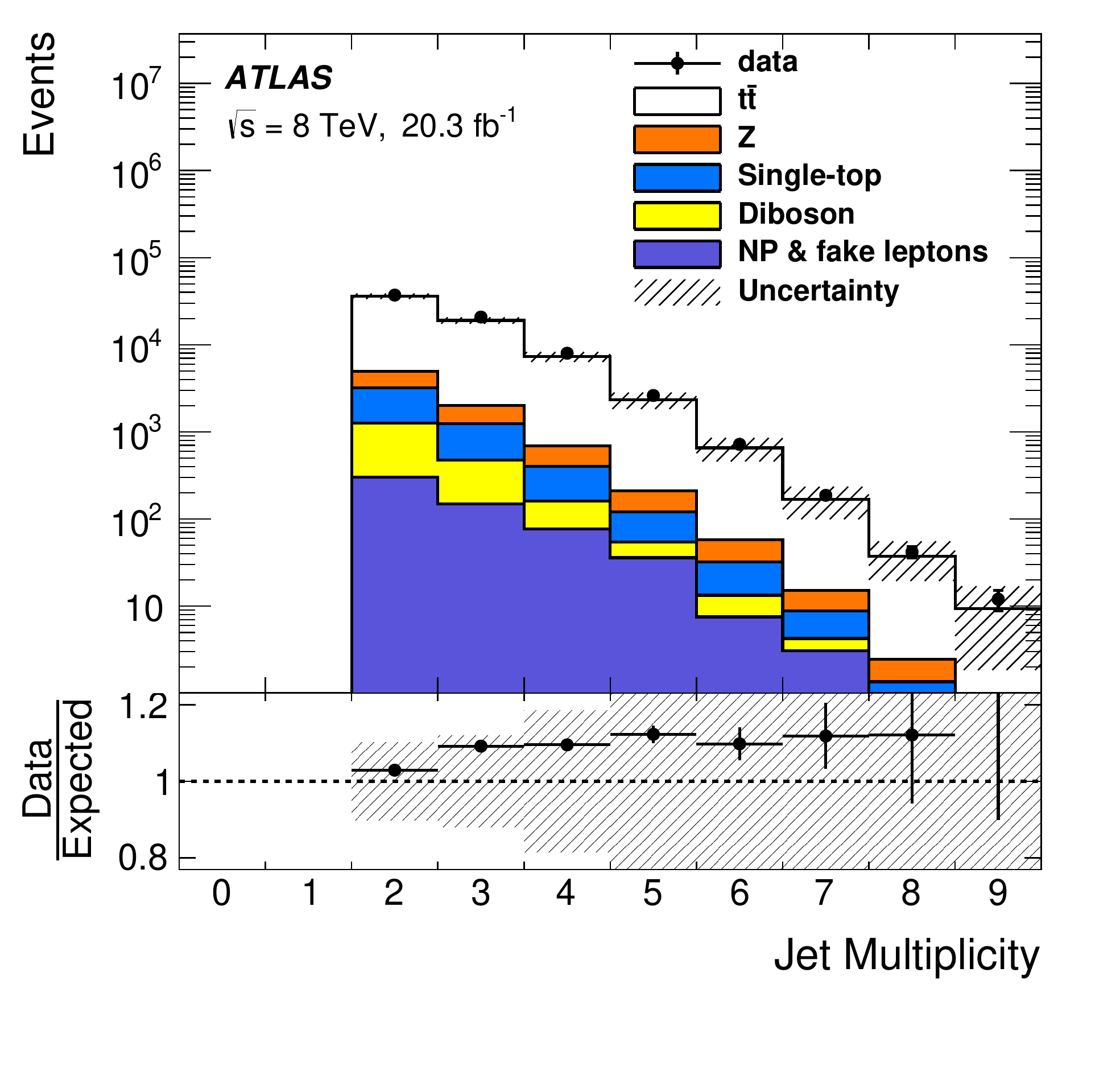}
   \includegraphics[width=0.49\textwidth]{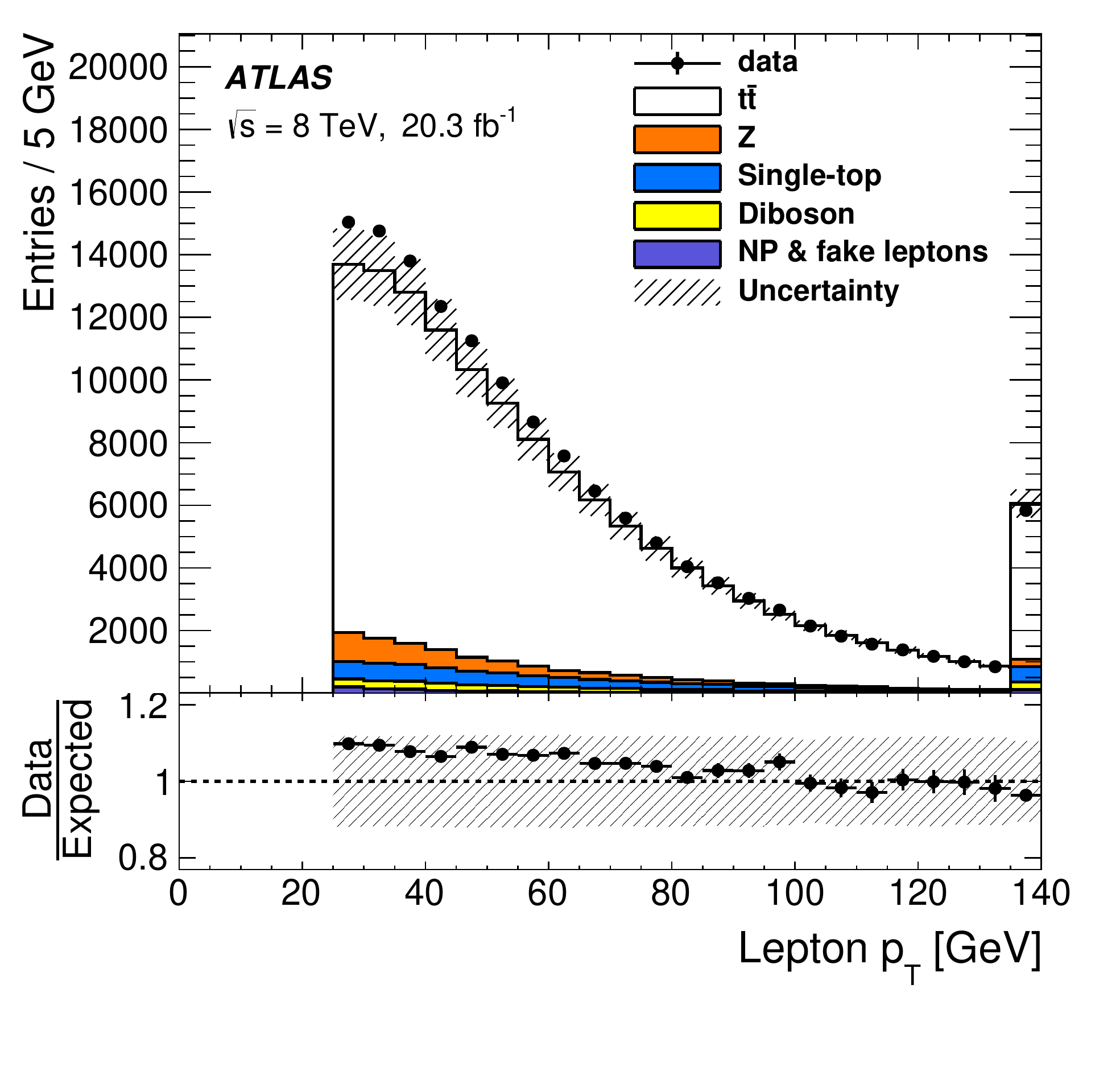}\\
   \includegraphics[width=0.49\textwidth]{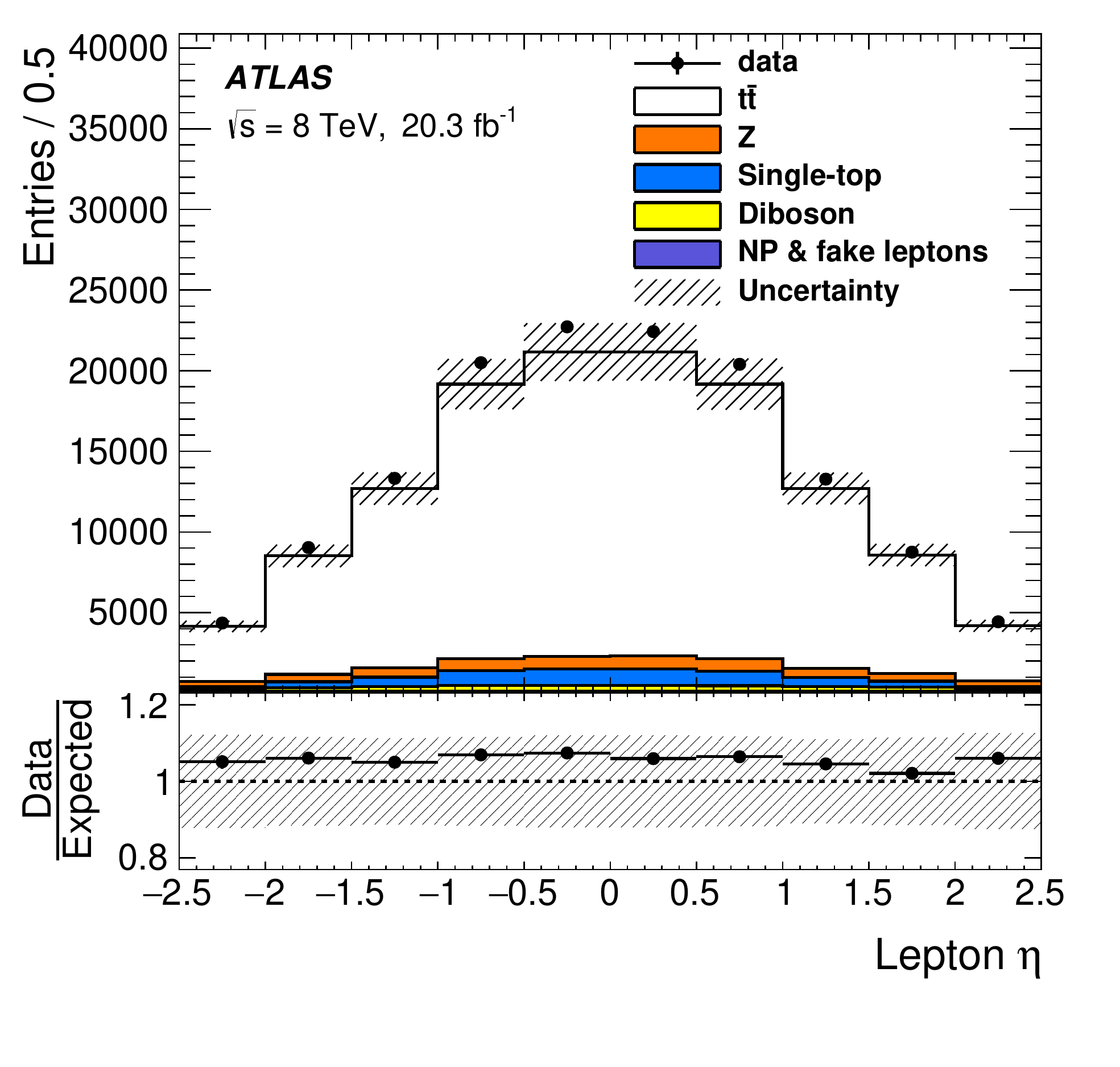}
  \caption{Distributions of the jet multiplicity, lepton $p_T$, and lepton
   $\eta$ for data (points) and predictions (histograms) for all
   channels combined after event selection. The data/expected
   ratio is also shown. The shaded area corresponds to the detector
   systematic uncertainty, the signal modeling systematic uncertainty, and the normalization uncertainty in signal and
   background. In the lepton $p_{\rm{T}}$ distribution, the last bin includes the overflow. }
  \label{fig:dataMC1}
 \end{center}
\end{figure}

The numbers of events for both expectation and data after
applying the selection criteria are shown in Table~\ref{tab:evtnumbers}
for the three final states. The uncertainties shown correspond to the total
uncertainty (including the statistical uncertainties from the limited
size of the MC simulated samples, as well as the systematic
uncertainties). The $e\mu$
channel contributes with the largest number of events, followed by
$\mu\mu$ and $ee$. Figure~\ref{fig:dataMC1} shows good agreement
within the systematic uncertainties between data and the predictions
as a function of jet multiplicity, lepton $p_T$ and $\eta$, for all
channels combined.
\begin{table}[htbp]
 \begin{center}
  \caption{Observed numbers of data events compared to the
   expected signal and background contributions in the three decay channels.
   The uncertainty corresponds to the total uncertainty in the given
  process. Data-driven (DD) scale factors are applied to the $Z+$jets
  and the NP \& fake leptons contributions. The
  $Z \rightarrow \tau\tau$ process in the $e\mu$ channel is estimated
  using MC simulation only. \label{tab:evtnumbers}
}
\renewcommand\arraystretch{1.1}
\centering
\resizebox{0.6\linewidth}{!}{
\sisetup{retain-explicit-plus}
\begin{tabular}{
 l
 r@{$\,\,$}
 c@{$\,\,$}
 r@{$\,\,\,\,\,\,$}
 r@{$\,\,$}
 c@{$\,\,$}
 r@{$\,\,\,\,\,\,$}
 r@{$\,\,$}
 c@{$\,\,$}
 r
}

\hline
\hline
Channel     & \multicolumn{3}{c}{ $ee$ }  & \multicolumn{3}{c}{$\mu\mu$ } & \multicolumn{3}{c} {$e\mu$}     \\
\hline

$t\bar{t}$          & 10200 & $\pm$ & 800 & 12100 & $\pm$ & 800 & 36000  & $\pm$ & 2400 \\
Single-top          & 510 & $\pm$ & 50 & 590 & $\pm$ & 50 & 1980  & $\pm$ & 170 \\
Diboson            & 31 & $\pm$ & 5 & 40 & $\pm$  & 6 & 1320  & $\pm$ & 100\\
$Z \rightarrow ee $ (DD)   & 1200 & $\pm$ & 260 & \multicolumn{3}{c}{--} & \multicolumn{3}{c}{--} \\
$Z \rightarrow \mu\mu $ (DD) & \multicolumn{3}{c}{--} & 1520 & $\pm$ & 300 & \multicolumn{3}{c}{--} \\
$Z \rightarrow \tau\tau $ (DD/MC)& 31 & $\pm$ & 15 & 58 & $\pm$ & 25 & 1120  & $\pm$ & 430 \\
NP \& fake leptons (DD) & 62 & & $^{\numRP{+119}{3}}_{\numRP{-29}{2}}$ & 45 & & $^{+36}_{-24}$ & 480 & &  $^{+240}_{-220}$ \\

Total Expected &  12000 & $\pm$ & 900 &  14400 & $\pm$ & 800 & 40900  & $\pm$ & 2500\\ \hline
Data & \multicolumn{3}{c}{12785} & \multicolumn{3}{c}{14453} &\multicolumn{3}{c}{42363} \\ \hline \hline

\end{tabular}}
 \end{center}
\end{table}

\section{Observables}
\label{sec:observables}
In dileptonic events, the charge asymmetry can be measured in two
 complementary ways: using the pseudorapidity of the charged leptons  or
using the rapidity of the top quarks.
The asymmetry based on the charged leptons uses the difference of the absolute pseudorapidity values of the positively and negatively charged leptons, $|\eta_{\ell^{+}}|$ and
$|\eta_{\ell^{-}}|$
\begin{linenomath}
\begin{equation}
\Delta |\eta| = |\eta_{\ell^{+}}|-|\eta_{\ell^{-}}|.
\end{equation}
\end{linenomath}
The leptonic asymmetry is defined as
\begin{linenomath}
\begin{equation}
\label{eq:ac_lep}
A_{\rm{C}}^{\ell\ell} = \frac{N(\Delta |\eta| >0) - N(\Delta |\eta| <0)}{N(\Delta |\eta|
  >0) + N(\Delta |\eta| <0)} ,
\end{equation}
\end{linenomath}
where $N(\Delta |\eta| >0)$ and $N(\Delta |\eta| <0)$ represent the number of events with
positive and negative $\Delta |\eta|$, respectively.
The SM prediction at NLO in QCD, including electroweak corrections, is
$\Acll = \Aclltheory$~\cite{Bernreuther:2012sx}, where the uncertainty
includes variations in scale and choice of PDF.
The leptonic asymmetry, that is slightly diluted with respect to the underlying top-quark asymmetry, has the advantage that no reconstruction of the
top--antitop quark system is required. Furthermore, it is also
sensitive to top-quark polarization effects, which occur in some
models predicting enhanced charge asymmetries.

For the \ttbar~charge asymmetry, the \ttbar~system has to be reconstructed and the absolute values of the  top and antitop quark rapidities ($|y_{t}|$ and $|y_{\bar{t}}|$, respectively) need to be computed.
Using
\begin{linenomath}
\begin{equation}
\Delta |y| = |y_{t}|-|y_{\bar{t}}|,
\end{equation}
\end{linenomath}
the $t\bar{t}$ charge asymmetry is defined as
\begin{linenomath}
\begin{equation}
\label{eq:ac}
A_{\rm{C}}^{t\bar{t}}= \frac{N(\Delta |y| > 0) - N(\Delta |y| < 0)}{N(\Delta |y|
  >0) + N(\Delta |y| <0)},
\end{equation}
\end{linenomath}
where $N(\Delta |y| > 0)$ and $N(\Delta |y| < 0)$ represent the number of events with
positive and negative $\Delta |y|$, respectively. The top (antitop) quarks are identified as those giving rise
to positive (negative) leptons.
The SM prediction at NLO QCD, including electroweak corrections, is
$\Ac = \Actheory$~\cite{Bernreuther:2012sx}.

The measurements of $\Acll$ and $\Ac$ are performed inclusively and
differentially as a function of $m_{t\bar{t}}$,  $p_{\rm{T},t\bar{t}}$, and
$\beta_{z,t\bar{t}}$. The fractions of quark--antiquark annihilation
and gluon fusion processes change as a function of
$m_{t\bar{t}}$, and thus an increasing asymmetry for increasing
$m_{t\bar{t}}$ is expected. Since $p_{\rm{T},t\bar{t}}$ depends on the
initial-state radiation, the asymmetry value is
expected to change as a function of $p_{\rm{T},t\bar{t}}$. In particular, the
contribution to the asymmetry from interference of diagrams with initial- and final-state radiation is negative, resulting in decreasing asymmetries
with increasing $p_{\rm{T},t\bar{t}}$. While the initial antiquark is
always a sea quark, the initial quark can be a valence quark. On
average, valence quarks have higher momenta than sea quarks, which can
result in a boost of the $t\bar{t}$ system in the direction of the
incoming quark. This results in an increased charge asymmetry for
increasing $\beta_{z,t\bar{t}}$. The asymmetry is also expected to be
different inclusively and differentially in different BSM models.


\section{Asymmetry measurements}
\label{sec:analysis}
The following measurements are performed:
\begin{itemize}
\item inclusive measurements of the  \ttbar and leptonic asymmetries, corrected for reconstruction and acceptance effects to parton level in the full phase space;
\item inclusive measurements of the  \ttbar and leptonic asymmetries, corrected for reconstruction effects to particle level in the fiducial region;
\item differential measurements of the \ttbar and leptonic asymmetries as a function of $m_{t\bar{t}}$, $p_{\rm{T},t\bar{t}}$, and $\beta_{z,t\bar{t}}$ in the fiducial region and the full phase space.
\end{itemize}

Particle-level results consider stable particles with a mean lifetime
larger than $0.3 \times 10^{-10}$~s. For the parton-level
measurements, MC generator-level objects are used. The
parton-level top quarks and leptons are selected after radiation.

The leptonic asymmetry can be extracted directly using the pseudorapidities of the measured charged leptons. For the \ttbar~charge asymmetry, the reconstruction of the top and antitop quark four-momenta is necessary. A kinematic method is used for the reconstruction, as described in Sec.~\ref{sec:reconstruction}.
Section~\ref{sec:fiducial} details the definition of the fiducial volume and the particle-level objects used for the fiducial measurement. In order to correct the measured asymmetry distributions  for detector and acceptance effects, an unfolding method, described in Sec.~\ref{sec:unfolding}, is used for all asymmetry measurements. Section~\ref{sec:measurement} describes how the various asymmetries are extracted.

\subsection{Top and antitop quark reconstruction} \label{sec:reconstruction}
For the reconstruction of the top and antitop quark four-momenta, a
kinematic reconstruction is used. The reconstruction is
performed by solving the system of equations that relate the particle
momenta at each of the decay vertices in the $t\bar{t} \rightarrow
W^{+}bW^{-}\bar{b} \rightarrow \ell^{+} \nu_{\ell} b \ell^{-} \bar{\nu_{\ell}} \bar{b} $  process.
Two neutrinos are produced and escape undetected. Thus, an
underconstrained system is obtained. This system is solved using the
kinematic (KIN) method~\cite{Abulencia:2006js,Aaltonen:2012tk},
assuming values of $172.5$~\GeV~and $80.4$~\GeV~for the top quark and $W$ boson masses, respectively, which
allows the system of equations to be solved numerically by the
Newton–Raphson method. 

If there are more than two reconstructed jets in a given event,
the two jets with the highest $b$-tagging weights (as determined by
the MV1 $b$-tagging algorithm) are used. This improves
the probability of choosing the correct jets, compared to just
choosing the two jets with the highest $p_{T}$,
from about 54\% to about 69\% in the inclusive selected sample. The
experimental uncertainties of the
measured objects (described in Sec.~\ref{sec:systematics}) are taken
into account by sampling the phase space of the measured jets and
$\MET$ according to their resolution in simulation.
The number of sampled points is called $N_{\rm{smear}}$, whose optimization is based on the time and efficiency of the top-pair reconstruction.
The resolution functions, obtained from the $t\bar{t}$ simulated sample, with respect to the jet $p_{\rm{T}}$ (for jets) and the
total transverse momentum in the event (for $\MET$) are used for the sampling.

For each sampling point, up to four solutions can be obtained.
The KIN method chooses the solution that leads to the lowest reconstructed mass of the $t\bar{t}$
system.  The reason for this is that the $t\bar{t}$ cross section is a
decreasing function of the partonic center-of-mass energy
$\sqrt{\hat{s}}\simeq m_{t\bar{t}}$, so events with smaller
$m_{t\bar{t}}$ are more likely.
There is also a twofold ambiguity in the lepton and $b$-jet
assignment. The correct assignment to the top and antitop quarks  is
chosen to be the one that has more reconstructed trials
$N_{\rm{smear}}^{\rm{reco}}$, i.e., the one that maximizes
$N_{\rm{smear}}^{\rm{reco}}/N_{\rm{smear}}$.
The chosen solution  is either the solution found using the nominal
jet energies and measured $\MET$, if available, or the first solution found during
the sampling.
The kinematic reconstruction fails for a given
event if no solution is found in any of the $N_{\rm{smear}}$ sampled
points. This is possible if, for example, the solution does not
converge within a given number of iterations.
The performance of the method is quantified by evaluating the
efficiency of reconstructing $t\bar{t}$ events that pass  dilepton
event selection, and the probability of reconstructing  the correct
sign of $\Delta |y|$. These probabilities are found to be  90\% and
76\%, respectively. The reconstruction efficiency is consistent
between data and the prediction.

\begin{figure}[!htbp]
 \begin{center}
 \includegraphics[width=0.49\textwidth]{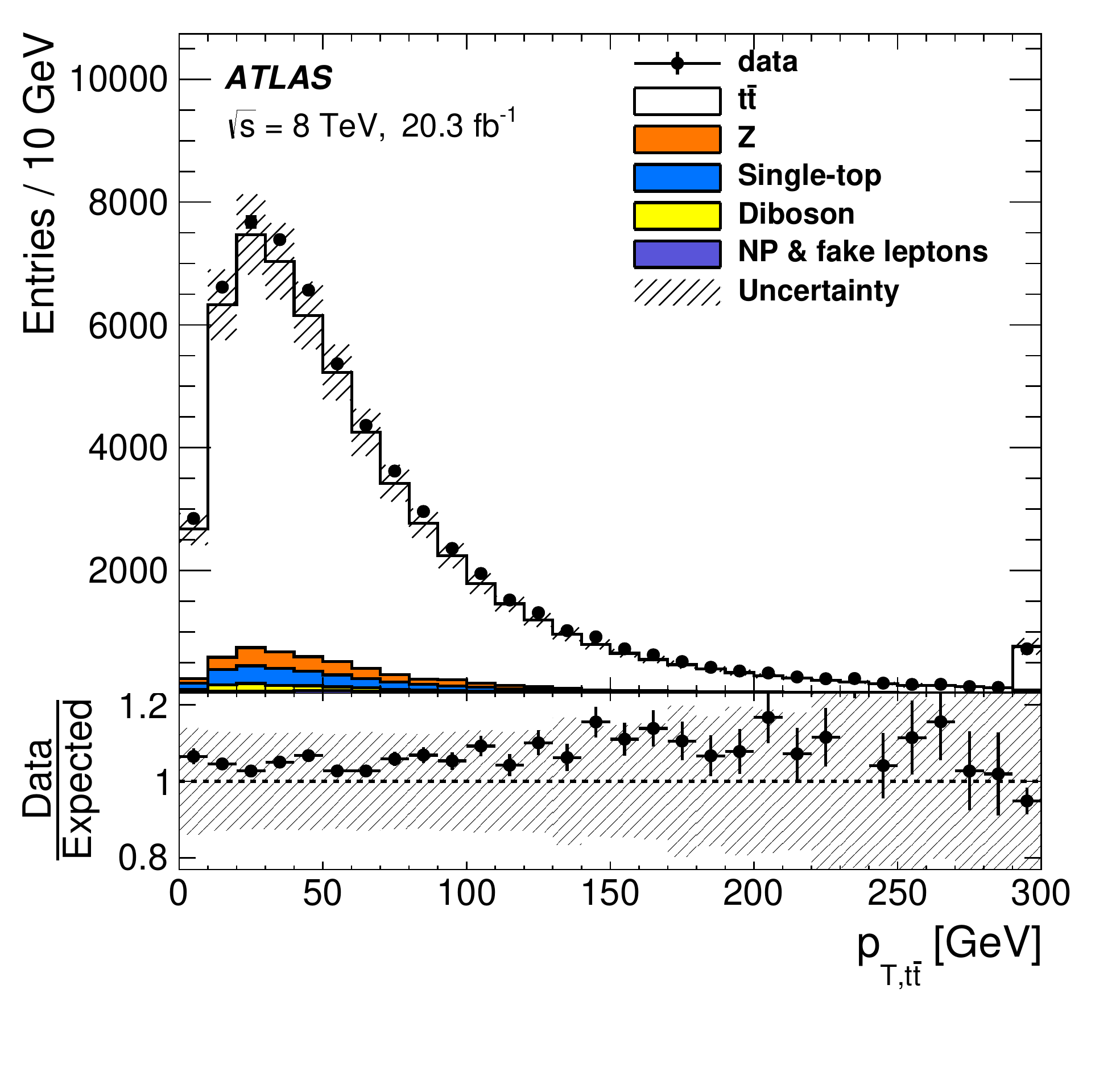}
 \includegraphics[width=0.49\textwidth]{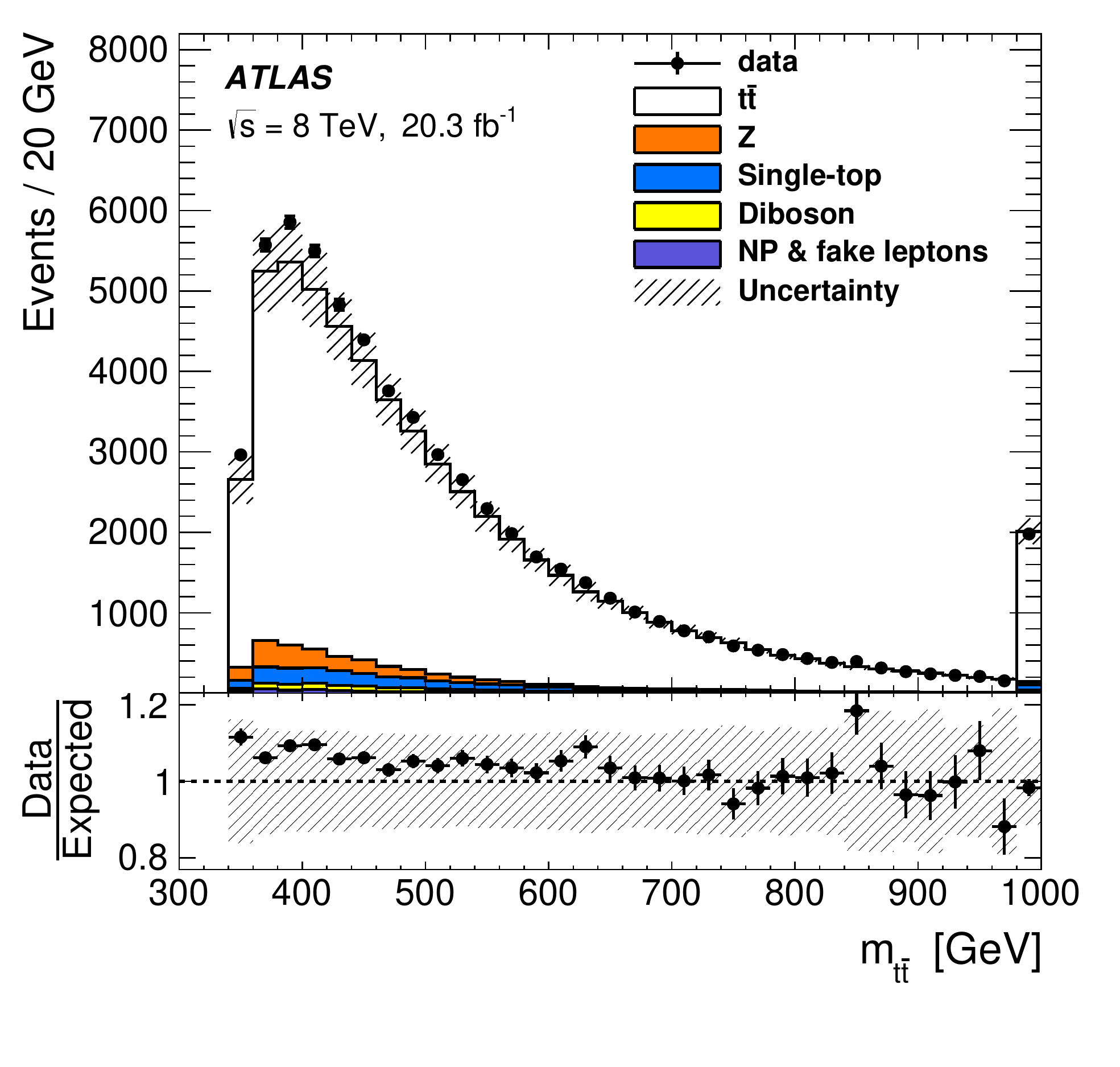}
 \includegraphics[width=0.49\textwidth]{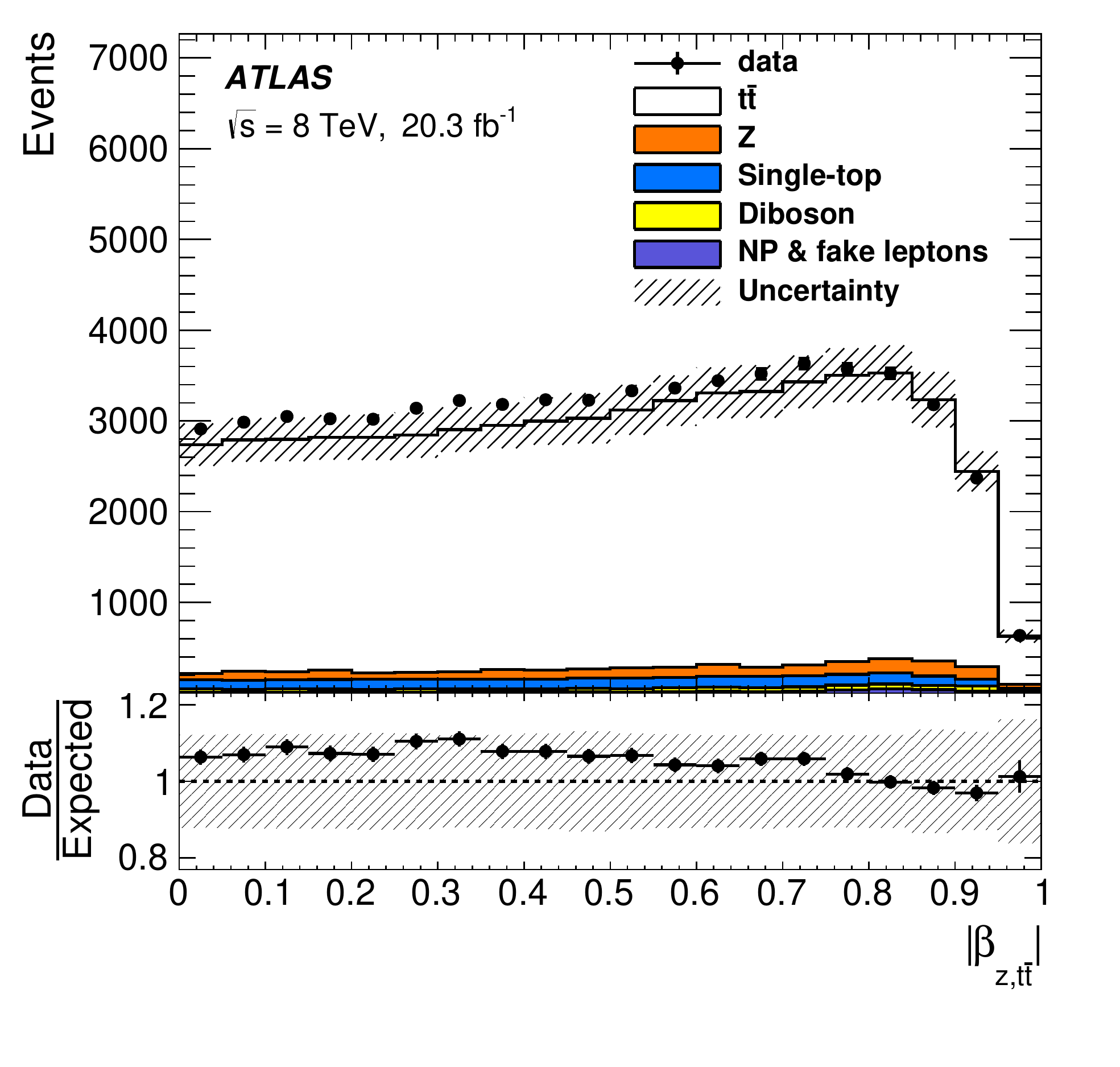}\\
   \caption{Distributions of $\pttt$, $\mtt$, and $|\betatt|$ for data (points) and predictions (histograms)
     after kinematic reconstruction. The data/expected ratio is also
     shown. The shaded area corresponds to the detector
   systematic uncertainty, the signal modeling systematic uncertainty, and the normalization uncertainty in signal and
   background. In the
      $\pttt$ and  $\mtt$  distributions, the last bin includes the
     overflow. }
   \label{fig:dataMC1reco}
 \end{center}
\end{figure}

Figure~\ref{fig:dataMC1reco} shows the distributions for data and prediction of the $p_T$, mass, and longitudinal boost of the $t\bar{t}$ system after applying the reconstruction method.  Good agreement between data and prediction is found.

\subsection{Particle-level objects and fiducial region} \label{sec:fiducial}
A fiducial region is defined in order to closely match the phase space
region accessed with the ATLAS detector and the requirements made on the reconstructed objects.
A fiducial measurement usually allows for MC generator dependencies to
be reduced, since it avoids large extrapolation to the full phase space.
In the fiducial region, only objects defined at particle level are used.

The considered charged leptons (electrons and muons) are required not to originate from hadrons.
Photons within $\Delta R = 0.1$ around the charged lepton are included in the four-momentum calculation.
The $\MET$ is
calculated as the summed four-momenta of neutrinos from the $W/Z$ boson decays, including those from $\tau$ decays.
Jets are reconstructed using the anti-$k_{t}$ algorithm with a radius
parameter $R = 0.4$. The electrons, muons,
neutrinos, and photons that are used in the definition of the selected
leptons are excluded from the clustering. Finally, identification of jets originating from $b$-quarks
is achieved using ghost matching~\cite{Cacciari:2008gn}.
The MC generator-level $b$-hadrons are clustered into the
particle-level jets, with their momenta scaled to a very small value. If a clustered
jet is found to contain a $b$-hadron, the particle-level jet is labelled as a
$b$-jet.

The fiducial volume is defined by requiring at least two particle-level jets and at least two leptons in the event,
both objects with $p_T >25$~\GeV~ and $|\eta| < 2.5$.
Events where leptons and jets overlap, within $\Delta R$  of $0.4$, are rejected. The particle-level jets are not required to be $b$-jets since this requirement is not shared between the three channels in the selection.

Using these objects, the reconstruction of top quarks (known as pseudotops~\cite{Aad:2015eia}) can be performed.
The assignment of the proper jet-lepton-neutrino permutation is chosen by first minimizing the difference between the mass computed from each lepton-neutrino combination and the $W$ boson mass value used in the MC simulation. Then, the difference between the mass of each combination of the chosen lepton-neutrino pairs with a jet and the top quark mass value, used in the MC simulation, is minimized. The $b$-jets are prioritized over the light jets for the proper jet-lepton-neutrino assignment.
The correlation coefficient between \deltay at the parton
and particle levels is found to be 79\%, while for \deltaeta it is 99\%.

The measurements of the asymmetry in the fiducial volume require the
treatment of an additional background contribution,
in which signal events from outside of the fiducial region migrate
into the detector acceptance due to resolution effects.
This nonfiducial background constitutes about 8\% of the expected
\ttbar events after selection,
as estimated by using MC simulation, and it was found to be independent
of the charge asymmetry value of the simulated sample.   A bin-by-bin
scale factor derived from simulation is applied to background-subtracted data to estimate the contribution of these events.

\subsection{Unfolding} \label{sec:unfolding}
The measurements are corrected for detector resolution and acceptance effects.
These corrections are performed using the fully Bayesian unfolding (FBU) technique~\cite{Fbu2012arXiv1201.4612C}.
The FBU procedure applies  Bayes' theorem to the problem of unfolding.
This application can be stated in the following terms: given an observed spectrum
$\Data$ with $N_r$ reconstructed bins and a migration matrix
$\TrasfMatrix$ with $N_r \times N_t$ bins giving the detector
response to a true spectrum with $N_t$ bins, the posterior probability
of the true spectrum $\Truth{}$ with $N_t$ bins follows the
probability density
\begin{linenomath}
\begin{equation}
\conditionalProb{\Truth{}}{\Data{}}
\propto{}
\conditionalLhood{\Data{}}{\Truth{}}
\cdot{}
\pi{}\left(\Truth{}\right),
\end{equation}
\end{linenomath}
where \conditionalLhood{\Data{}}{\Truth{}} is the
 likelihood of \Data{} assuming \Truth{} and \TrasfMatrix{}, and $\pi{}$ is
the prior probability density for the true spectrum \Truth{}.
The selection and reconstruction efficiency, which is the probability
that an event produced in MC generator-level bin $t$ is reconstructed in
one of the $N_r$ bins included in \TrasfMatrix{}, is taken into
account in the likelihood.
An uninformative
 prior probability density is chosen, such that equal probabilities
 are assigned to all \Truth{} spectra within a wide range. 
  The background in each bin is taken into account when computing
\conditionalLhood{\Data{}}{\Truth{}}. The unfolded spectrum and its
associated uncertainty are extracted from the posterior probability
density distribution.

The migration matrix is obtained from the nominal
$t\bar{t}$ simulated sample using the top quarks
before their decay (parton level) or pseudotops (particle
level). The combination of the three decay channels is performed by
using a rectangular migration matrix, which maps the reconstructed
distribution of the three channels to the same corrected
distribution.

To validate the method, a linearity test is performed for the inclusive and differential measurements of the charge asymmetry. A given
asymmetry value is introduced by reweighting the samples according to a nonlinear function of
$\Delta|y|$ and $\Delta|\eta|$ based on a BSM axigluon model~\cite{AguilarSaavedra:2009es}. The asymmetry values are in the range of  $-6\%$ to $6\%$ in steps of $2\%$. Good agreement between the unfolded values and the injected values is found, and the calibration curves
derived from this test are linear.

For the treatment of systematic uncertainties in the Bayesian inference approach,
the likelihood \conditionalLhood{\Data{}}{\Truth{}} is extended
with nuisance parameter terms. This marginal likelihood is defined as
\begin{linenomath}
\begin{equation}
\conditionalLhood{\Data{}}{\Truth{}} = \int
\conditionalLhood{\Data{}}{\Truth{},\vect{\theta}}\cdot \pi(\vect{\theta}) d\vect{\theta} \, ,
\end{equation}
\end{linenomath}
where $\vect{\theta}$ are the nuisance parameters, and $\pi(\vect{\theta})$ their prior probability densities,
which are assumed to be normal distributions $\mathcal{N}$ with a mean
value of zero and a variance of one.
A nuisance parameter is associated with each of the uncertainty sources.
As is described in Sec.~\ref{sec:systematics}, four categories of uncertainties are considered in this analysis,
but only two are included in the marginalization:
the normalizations of the background processes ($\vect{\theta}_b$),
and the uncertainties associated with the object identification,
reconstruction and calibration ($\vect{\theta}_s$).
While the first ones only affect the background predictions, the latter, referred to as
object systematic uncertainties, affect both the reconstructed distribution for the $t\bar{t}$ signal  ($\vect{R}(\Truth{}; \vect{\theta}_s)$) and the total background prediction ($\vect{B}(\vect{\theta}_s, \vect{\theta}_b)$). The marginal likelihood then becomes
\begin{linenomath}
\begin{equation}
\conditionalLhood{\Data{}}{\Truth{}} = \int
\conditionalLhood{\Data{}}{\vect{R}(\Truth{}; \vect{\theta}_s),\vect{B}(\vect{\theta}_s, \vect{\theta}_b)} \cdot
\mathcal{N}(\vect{\theta}_s) \cdot \mathcal{N}(\vect{\theta}_b)\ d\vect{\theta}_s\ d\vect{\theta}_b \ .
\end{equation}
\end{linenomath}

\subsection{Binning optimization and asymmetry extraction} \label{sec:measurement}

For each measurement, the choice of binning for the $\Delta |y|$ and $\Delta|\eta|$ distributions is optimized
by minimizing the expected statistical uncertainty while allowing only a negligible bias in the linearity of the calibration curve.
The optimal binnings are found to be $4$ and $16$ bins in an interval between $-5$ and $5$
  for the inclusive measurements  of the $\Delta |y|$ and
  $\Delta|\eta|$ distributions, respectively. For the differential
  measurements, $4$ bins are used for the $\Delta |y|$ and
  $\Delta|\eta|$ distributions for each of the chosen
$m_{t\bar{t}}$, $p_{\rm{T},t\bar{t}}$ and $\beta_{z,t\bar{t}}$ ranges.
Due to the limited size of the data sample, only two ranges of values are
considered for the  $m_{t\bar{t}}$, $p_{\rm{T},t\bar{t}}$ and
$\beta_{z,t\bar{t}}$ variables.
The charge asymmetry predicted in the SM is expected to increase as a function of
$m_{t\bar{t}}$ while it is expected to be large for low
$p_{\rm{T},t\bar{t}}$ and small and roughly constant for higher $p_{\rm{T},t\bar{t}}$.
The exact boundary between the bins for $m_{t\bar{t}}$ was chosen to minimize
the expected uncertainties in the bins. For $p_{\rm{T},t\bar{t}}$, the boundary was
set at 30 \GeV as a compromise between the uncertainty optimization and
the interest in the $p_{\rm{T},t\bar{t}}$ dependence described above. For $\beta_{z,t\bar{t}}$,
the boundary at 0.6 is motivated by the large difference of the predicted
asymmetry between SM and BSM models in the range (0.6,1.0)~\cite{atlas8tevljets}.
Table~\ref{tab:binning} summarizes the differential bins used in the analysis.

\begin{table}[h]
\renewcommand\arraystretch{1.4}
\centering
\caption{Bins and ranges used for the inclusive and differential
  measurements. The binning choices used in the $\Delta|\eta|$ and $\Delta|y|$
  distributions are shown. The bins are symmetric around zero.}
\label{tab:binning}
\begin{tabular}{cccc}
\hline
\hline

& & $\Delta|\eta|$ & $\Delta|y|$   \\ \hline

\multicolumn{2}{c}{Inclusive}   & $[0.0, 0.3, 0.6, 0.9, 1.2, 1.5, 1.7, 1.9, 5.0]$ & $[0.0,0.75,5.0]$  \\ \hline
\multirow{2}{*}{$m_{t\bar{t}}$} & 0--500 \GeV & $[0.0,0.8,5.0]$ & $[0.0,0.6,5.0]$  \\
& 500--2000 \GeV & $[0.0,1.4,5.0]$ & $[0.0,1.2,5.0]$  \\ \hline
\multirow{2}{*}{$\beta_{t\bar{t}}$} &  0--0.6  & $[0.0,0.8,5.0]$ & $[0.0,0.5,5.0]$  \\
 &  0.6--1.0 & $[0.0,1.2,5.0]$ & $[0.0,0.9,5.0]$  \\ \hline
\multirow{2}{*}{$p^{t\bar{t}}_{\rm{T}}$} &   0--30 \GeV & $[0.0,0.7,5.0]$ & $[0.0,0.8,5.0]$  \\
 &  30--1000 \GeV & $[0.0,0.7,5.0]$ & $[0.0,0.8,5.0]$  \\ \hline
\hline
 \end{tabular}
\end{table}

\begin{figure}[h]
  \begin{center}
    \includegraphics[width=0.95\columnwidth]{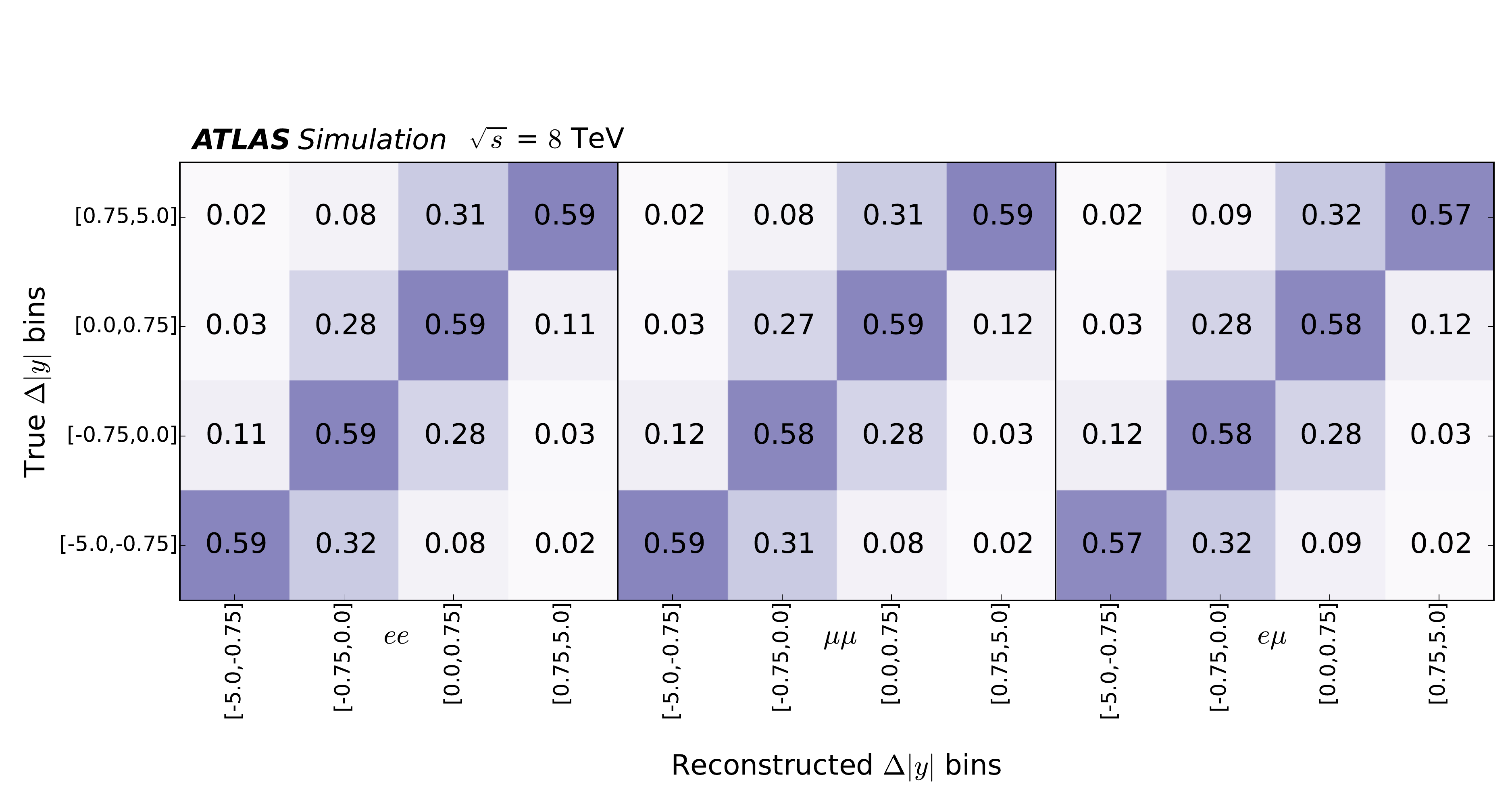}
  \end{center}
 \caption{Rectangular migration matrix for the $\Delta |y|$
          observable in the fiducial volume. The first four columns correspond to the $ee$ channel, followed by $\mu\mu$ and $e\mu$.
          The numbers are normalized by row for each channel. }
\label{fig:ttasym_responsemat}
\end{figure}

\begin{figure}[h]
  \begin{center}
    \includegraphics[width=0.95\columnwidth]{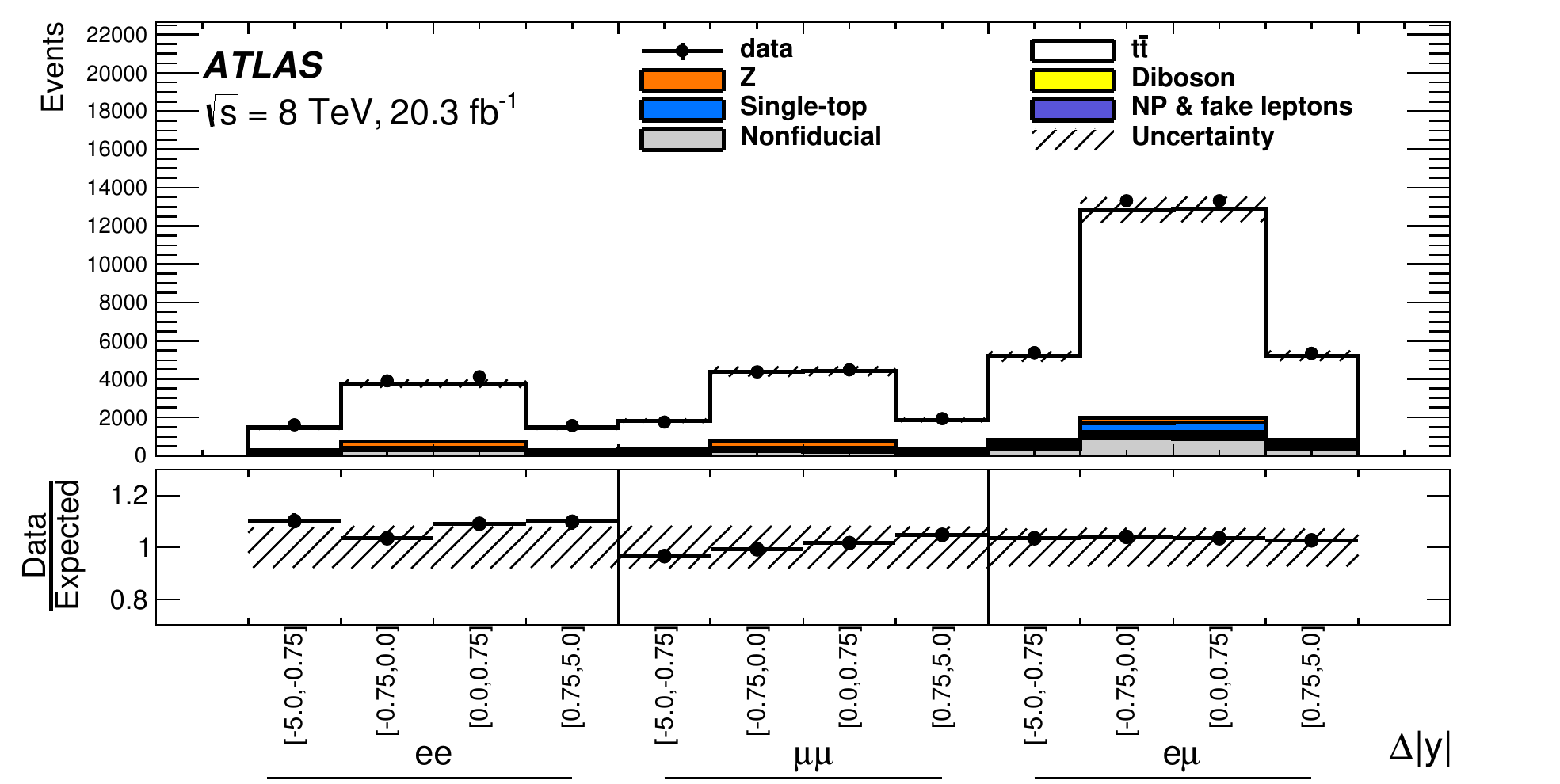}\\
    \includegraphics[width=0.95\columnwidth]{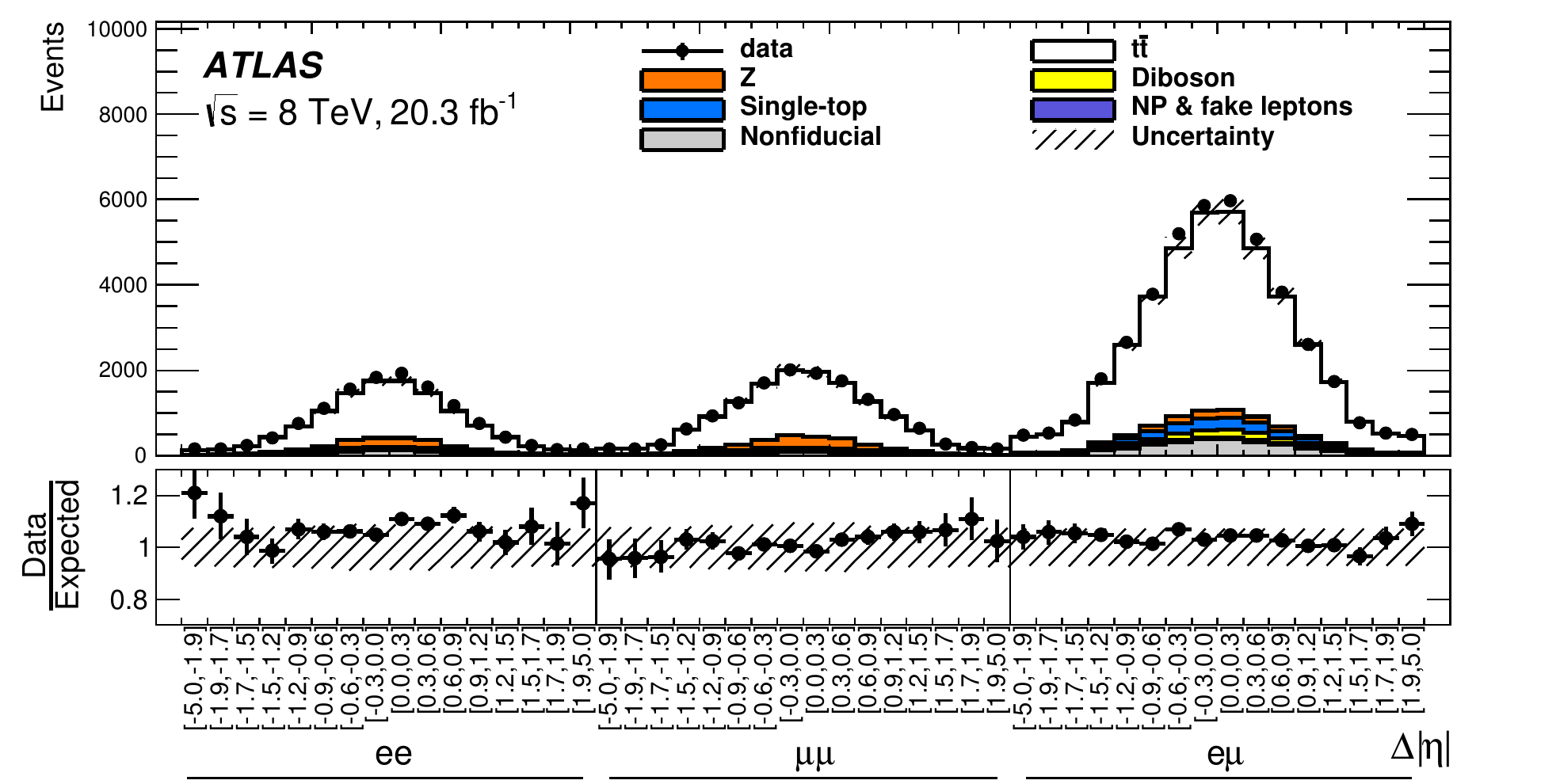}
  \end{center}
 \caption{Input distributions for the inclusive $t\bar{t}$ (top) and leptonic (bottom) asymmetry measurements. The first 4 $\Delta|y|$ and 16 $\Delta|\eta|$  bins correspond to the $ee$ channel, followed by $\mu\mu$ and $e\mu$. The bin boundaries are symmetric around zero and are defined as $[0.0,0.75,5.0]$ and $[0.0, 0.3, 0.6, 0.9, 1.2, 1.5, 1.7, 1.9, 5.0]$ for $\Delta|y|$  and $\Delta|\eta|$, respectively. The data/expected ratio is also shown. }
\label{fig:ttasym_inputdist}
\end{figure}

For the optimized binning choice, more than $50 \%$ of the events
populate  the diagonal bins of the migration matrix for the
$\Delta|y|$ distribution, and more than $97\%$ for $\Delta|\eta|$. The
rectangular migration matrix, normalized by row for each channel, used for the
inclusive $t\bar{t}$ asymmetry measurement is shown in
Fig.~\ref{fig:ttasym_responsemat}.
Due to the nonuniform shape of the  $\Delta|y|$  distribution, the
matrix is not symmetric around the diagonal.
 The migrations are symmetric around zero
and do not affect the asymmetry value. The  $\Delta|y|$ and
$\Delta|\eta|$ input distributions used for the inclusive measurement
are shown in Fig.~\ref{fig:ttasym_inputdist}.

The asymmetry values are extracted by taking the mean of the posterior probability density obtained during the unfolding procedure. The uncertainty is obtained from the standard deviation of the posterior probability density.

\section{Systematic uncertainties}
\label{sec:systematics}
Four classes of systematic uncertainties affect the measurement of the
charge asymmetry:  detector modeling uncertainties, uncertainties
related to the estimation of the backgrounds, signal modeling
uncertainties, and other uncertainties, which involve the top-quark
reconstruction, the bias introduced by the unfolding procedure and the
MC statistical uncertainty.

The first two categories are estimated within the unfolding
through the marginalization procedure where the total uncertainty
includes these systematic uncertainties together with the
statistical uncertainty.
In order to estimate the impact of each source of systematic uncertainty, pseudodata corresponding to the sum of the nominal signal and background samples is used. The unfolding procedure with marginalization is applied to the pseudodata and constraints on the systematic uncertainties are obtained. These constraints are then used to build the $\pm 1\sigma$ variations of the prediction. The varied pseudodata are then unfolded without marginalization. The impact of each systematic uncertainty is computed by taking half of the difference between the results obtained from the $\pm 1\sigma$ variations of pseudodata. Clearly, this is only
 an approximate estimate of the individual contribution of
each source of systematic uncertainty within the overall marginalization procedure.

The signal modeling uncertainties are not estimated through the marginalization
procedure. For these uncertainties, the migration matrix is fixed to
the nominal $t\bar{t}$ sample and
distributions  obtained with different generators and different
injected asymmetries are unfolded. The unfolded asymmetries are
compared with the injected asymmetries and the calibration curves are
obtained.
The slopes and offsets of the calibration curves are extrapolated to
the measured value in data.

The final category of systematic uncertainties involves different estimation
methods. The uncertainty related to the top-quark reconstruction is
estimated on pseudodata by varying the starting point of the smearing
procedure within the kinematic reconstruction and repeating the
unfolding. The bias introduced by the unfolding procedure is estimated
by propagating the
residual slope and offset of the nominal calibration curve to
the measured value. The MC statistical uncertainty is estimated by
varying the nominal migration matrix within the MC statistical
uncertainty and the unfolding procedure is repeated for each
variation.
All sources of systematic uncertainties are discussed below in detail.

\subsection{Detector modeling uncertainties}

\subsubsection*{Lepton-related uncertainties}

The reconstruction and identification efficiencies of electrons and muons, as well as the efficiency of the
triggers used to record the events, differ between data and simulation. Scale factors, and their uncertainties, are derived using
tag-and-probe techniques on $Z \rightarrow \ell^{+} \ell^{-} (\ell = e,\,\mu)$ in data and in simulated samples to correct the simulation for these differences~\cite{Aad:2012xs,Aad:2014fxa,Aad:2014rra,Aad:2014nim}. Moreover, the accuracy of the
lepton momentum scale and resolution in simulation is also checked using
reconstructed distributions of the  $Z \rightarrow \ell^{+} \ell^{-}$
and $J/\psi \rightarrow  \ell^{+} \ell^{-} $ masses. In the case of
electrons, $E/p$ studies using $W \rightarrow e\nu$ events are also
used. Small differences are observed between data and
simulation. Corrections for the lepton energy scale and resolution,
and their related uncertainties are
considered~\cite{,Aad:2014fxa,Aad:2014rra,Aad:2014nim}.
The uncertainties are propagated through this analysis and represent a minor source of uncertainty in the measurements.

\subsubsection*{Jet-related uncertainties}
The jet energy scale and its uncertainty are derived combining information from test-beam
data, LHC collision data, and simulation~\cite{Aad:2014bia}. The jet energy scale uncertainty is split
into 22 uncorrelated sources that have different jet $p_{\rm{T}}$ and $\eta$ dependencies
and are treated independently in this analysis.
The total jet energy scale uncertainty is one of the dominant
uncertainties in $\Actt$ and in the differential measurements of $\Acll$.  The jet reconstruction efficiency is found to be about $0.2\%$ lower in simulation than in data for jets
below $30$~\GeV~and consistent with data for higher jet $p_T$. All jet-related
kinematic variables (including the missing transverse momentum) are recomputed by removing randomly $0.2\%$ of the jets with
$p_{\rm{T}}$ below $30$~\GeV~and the event selection is repeated. The efficiency for each jet to satisfy the JVF requirement is measured in $Z \rightarrow \ell^{+} \ell^{-}+\textrm{1-jet}$ events
in data and simulation~\cite{Aad:2015ina}. The corresponding uncertainty is evaluated in the analysis by changing the nominal JVF cut value and
repeating the analysis using the modified cut value. The uncertainty related to the jet energy resolution is estimated
by smearing the energy of jets in simulation by the  difference between the jet energy resolutions for data and simulation~\cite{Aad:2012ag}. Finally, the efficiencies to tag jets from $b$- and $c$-quarks, light quarks, and gluons in simulation are
corrected by $p_T$- and $\eta$-dependent data/MC scale factors~\cite{Aad:2015ydr,ATLAS-CONF-2014-004,ATLAS-CONF-2014-046,}.
The uncertainties in these scale factors are propagated to the measured value. The impact on the measurement of the jet reconstruction efficiency, jet vertex
fraction, jet resolution, and jet tagging efficiency is minor.

\subsubsection*{Missing transverse momentum}

The systematic uncertainties associated with the momenta and energies
of reconstructed objects (leptons and jets) are also propagated to the $\met$ calculation. The $\met$ reconstruction
also receives contributions from the presence of low-$p_{\rm{T}}$ jets and calorimeter cells not included in reconstructed objects (``soft terms''). The systematic uncertainty of the soft terms is evaluated using
$Z \rightarrow \mu^{+} \mu^{-}$ events using methods similar to those used in Ref.~\cite{Aad:2012re}.
The uncertainty has a negligible effect on the measured asymmetries.

\subsection{Background-related uncertainties}
\label{sec:syst_background}

The uncertainties in the single-top-quark and diboson backgrounds are about 7\% and 5\%, respectively.
These correspond to the uncertainties in the theoretical cross sections used for the normalization of the MC simulated samples.

The uncertainty in the normalization of the fake-lepton background is
evaluated by using various Monte Carlo simulations for each process
contributing to this background and propagating the change into the number
of expected events in the signal region.
In the $\mu\mu$ channel, the uncertainty is obtained by comparing a
purely data-driven method based on the measurement of the
efficiencies for real and fake loose leptons,
 and the estimation used in this analysis.
Following a Bayesian procedure assuming constant
a priori probability for a non-negative number of events, the
resulting total relative uncertainties are
 $^{+193\%}_{-47\%}$ in the $ee$, $^{+80\%}_{-53\%}$ in the $\mu\mu$, and $^{+49\%}_{-45\%}$ in the
$e\mu$ channel, where the uncertainties correspond to the $68$\% central probability region.

In the case of the Drell--Yan events, the detector modeling
systematic uncertainties described previously are propagated to the scale factors
derived in the control region by recalculating them for all the systematic uncertainty variations.
An additional uncertainty of 6\% is estimated by varying the $Z$ mass window of the control
region used to obtain the scale factors and is added in quadrature to
obtain the final uncertainty in these scale factors.

This category represents a minor source of uncertainty in the measurement.

\subsection{Signal modeling uncertainties}
The uncertainty due to the choice of MC generator is obtained by taking the full difference between the \powheg
and \mcatnlo predictions, both interfaced with \herwig, while the uncertainty from parton showering and hadronization is obtained by
comparing \powheg interfaced with either \pythia or \herwig.
These components are among the dominant uncertainties.
The effect produced by the different amount of ISR and FSR in the
events is estimated as half the difference between the asymmetries
obtained from MC samples with more or less ISR/FSR.
These samples are generated with \powheg interfaced with \pythia
for which the parameters of the generation were varied to span the
ranges compatible with the results
of measurements of $t\bar{t}$ production in association with
jets~\cite{TOPQ-2011-21}.
Finally, PDF uncertainties are obtained by
using the error sets of CT10,
MWST2008 and NNPDF2.3,
and following the prescriptions recommended by the PDF4LHC working
group~\cite{Botje:2011sn}. The impact of the last two uncertainties is small.

\subsection{Other uncertainties}

\subsubsection*{Top-quark kinematic reconstruction}
There is an intrinsic uncertainty of the reconstruction method due to
the randomness in the smearing procedure.
If the smearing starts from a different point it
could lead to a different solution. The uncertainty from this effect
is computed by performing pseudoexperiments on MC events. For each
event, the $t\bar{t}$ system is reconstructed multiple times varying
the starting point of the smearing procedure. Then, for each variation
the unfolding procedure is repeated and the standard deviation of the
asymmetries obtained is taken as the uncertainty. This represents one
of the major systematic uncertainties for the measurements, but it is
still only half of the statistical uncertainty for most of them.

\subsubsection*{Nonclosure uncertainties}
When the calibration curve for the nominal signal \powheg sample is estimated a residual slope and a nonzero offset are observed. This bias, introduced by the unfolding procedure, is propagated to the measured values in the same way as for the signal modeling uncertainties. This source of uncertainty is negligible in all the measurements.

\subsubsection*{MC sample size}
The uncertainty associated with the limited size of the nominal signal \powheg sample is evaluated by performing pseudoexperiments on MC events. The migration matrix is varied within the MC statistical uncertainty and the unfolding procedure is repeated. The standard deviation of the obtained asymmetries is taken as the uncertainty. This uncertainty has a minor impact on the measurements.

\subsection{Summary of systematic uncertainties}
Tables \ref{tab:uncertainties_lepton} and \ref{tab:uncertainties_ttbar} show how each category of uncertainty affects the measurements of the lepton and \ttbar\ asymmetry, respectively. The statistical uncertainty gives the largest contribution to the measurement, followed by the reconstruction and the signal modeling uncertainties. The signal modeling uncertainties are enhanced in the differential measurements by the migrations between the differential bins across the different MC generators used for their estimation. The uncertainty obtained by the sum in quadrature of the individual systematic uncertainties is slightly larger than the total marginalized uncertainty in the measurements.

\begin{table}[h]
\renewcommand\arraystretch{1.4}
\centering
\caption{Absolute uncertainties from the different sources affecting the leptonic asymmetry of the three channels combined in the fiducial and full phase space.}
\label{tab:uncertainties_lepton}
\resizebox{\linewidth}{!}{\begin{tabular}{cccccccp{0.5cm}ccccc}
\hline
\hline

& & \multicolumn{11}{c}{Absolute uncertainties in $A_{\rm{C}}^{\ell \ell}$ }   \\ \hline

& & \multicolumn{5}{c}{Fiducial volume} & & \multicolumn{5}{c}{Full phase space} \\ \hline
& & Statistics & Detector & Bkg & Signal modeling & Other & & Statistics & Detector & Bkg & Signal modeling & Other \\ \hline \hline

 \multicolumn{2}{c}{Inclusive}   & $0.005$ & $0.001$ & $0.001$ & $0.002$ &  $0.001$ & & $0.005$ & $0.001$ & $0.001$ & $0.004$  & $0.001$  \\ \hline
\multirow{2}{*}{$m_{t\bar{t}}$} & 0--500 \GeV & $0.008$ & $0.002$ & $0.001$ & $0.005$ &  $0.005$& & $0.008$ & $0.002$ & $0.001$ & $0.005$  & $0.006$ \\
& 500--2000 \GeV & $0.012$ & $0.004$ & $<0.001$ & $0.013$  & $0.005$ && $0.011$  & $0.004$ & $<0.001$ & $0.014$  & $0.005$  \\ \hline
\multirow{2}{*}{$\beta_{t\bar{t}}$} &  0--0.6  & $0.007$ & $0.003$ & $<0.001$ & $0.004$ & $0.004$ & & $0.007$ & $0.002$ & $<0.001$ & $0.005$  & $0.005$  \\
 &  0.6--1.0 & $0.010$ & $0.005$ & $0.001$ & $0.005$  & $0.004$ & & $0.010$ & $0.003$ & $0.001$ & $0.006$  & $0.004$  \\ \hline
\multirow{2}{*}{$p^{t\bar{t}}_{\rm{T}}$} &   0--30 \GeV & $0.015$  & $0.009$ & $0.001$ & $0.015$  & $0.006$ & & $0.015$  & $0.010$ & $0.001$ & $0.017$  & $0.007$ \\
 &  30--1000 \GeV & $0.011$  & $0.004$ & $0.001$ & $0.012$  & $0.005$& & $0.010$  & $0.004$ & $0.001$ & $0.013$  & $0.006$ \\ \hline
\hline

 \end{tabular}}

\end{table}

\begin{table}[h]
\renewcommand\arraystretch{1.4}
\centering
\caption{Absolute uncertainties from the different sources affecting the $t\bar{t}$ asymmetry of the three channels combined in the fiducial and full phase space.}
\label{tab:uncertainties_ttbar}
\resizebox{\linewidth}{!}{\begin{tabular}{cccccccp{0.5cm}ccccc}
\hline
\hline

& & \multicolumn{10}{c}{Absolute uncertainties in $A_{\rm{C}}^{t\bar{t}}$ }   \\ \hline

& & \multicolumn{5}{c}{Fiducial volume}  & & \multicolumn{5}{c}{Full phase space} \\ \hline
& & Statistics & Detector & Bkg & Signal modeling & Other & & Statistics & Detector & Bkg & Signal modeling & Other \\ \hline \hline

 \multicolumn{2}{c}{Inclusive}   & $0.013$ & $0.008$ & $<0.001$ & $0.007$  & $0.007$ &  &$0.011$ &  $0.006$ & $<0.001$ & $0.008$  & $0.006$  \\ \hline
\multirow{2}{*}{$m_{t\bar{t}}$} & 0--500 \GeV & $0.030$ & $0.024$ & $0.001$ & $0.016$  & $0.021$ &  &$0.028$&  $0.021$ & $0.002$ & $0.018$  & $0.020$ \\
& 500--2000 \GeV & $0.018$ & $0.007$ & $<0.001$ & $0.015$  & $0.009$ &  &$0.015$ & $0.006$ & $<0.001$ & $0.016$  & $0.008$  \\ \hline
\multirow{2}{*}{$\beta_{t\bar{t}}$} &  0--0.6  & $0.023$ & $0.021$ & $0.002$ & $0.014$ & $0.018$ &  & $0.023$ & $0.019$ & $0.002$ & $0.015$  & $0.017$  \\
 &  0.6--1.0 & $0.021$ & $0.009$ & $0.001$ & $0.013$  & $0.011$ &  &$0.018$  & $0.009$ & $0.001$ & $0.013$  & $0.010$  \\ \hline
\multirow{2}{*}{$p^{t\bar{t}}_{\rm{T}}$} &   0--30 \GeV & $0.035$  & $0.019$ & $0.003$ & $0.018$  & $0.020$ & & $0.031$ & $0.015$ & $0.004$ & $0.019$  & $0.017$ \\
 &  30--1000 \GeV & $0.027$  & $0.015$ & $0.003$ & $0.018$  & $0.017$& &$0.025$  & $0.013$ & $0.003$ & $0.014$  & $0.015$ \\ \hline
\hline

 \end{tabular}}

\end{table}

\section{Results}
\label{sec:result}
\begin{figure}[h]
\begin{center}
\includegraphics[width=0.70\columnwidth]{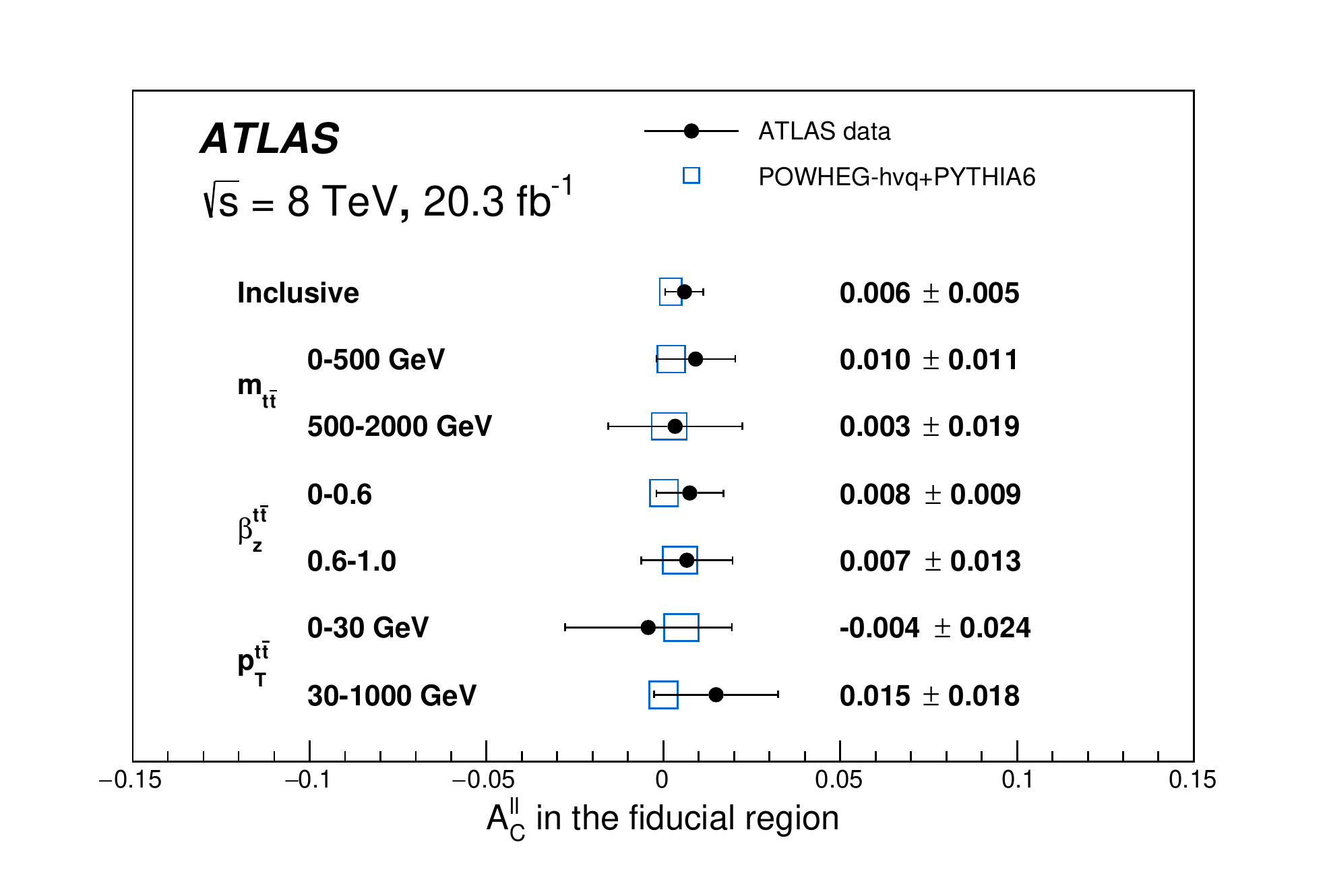}\\
\includegraphics[width=0.70\columnwidth]{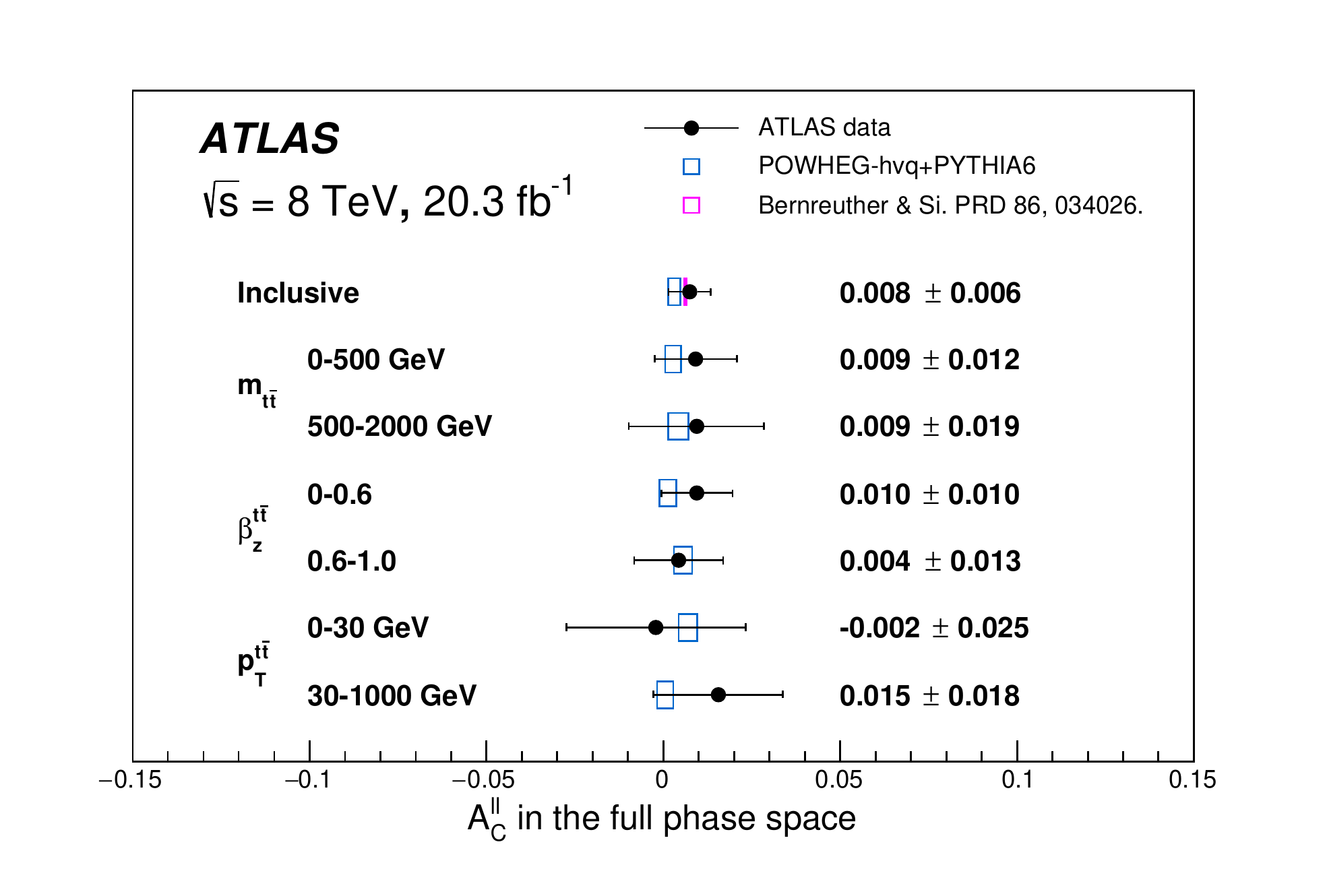}
\caption{Summary of all the measurements in this paper for the leptonic
asymmetry in the fiducial volume (top) and full phase space
(bottom). The predictions shown in blue are obtained using \powheg
+ \pythia at NLO where the uncertainties are statistical, and the
corresponding theoretical uncertainties are
small compared to the experimental precision. The inclusive
measurement in the full phase space is compared to a NLO + EW
prediction ~\cite{Bernreuther:2012sx}.}
\label{fig:sumary_plot_lepton}
\end{center}
\end{figure}

\begin{figure}[h]
\begin{center}
\includegraphics[width=0.70\columnwidth]{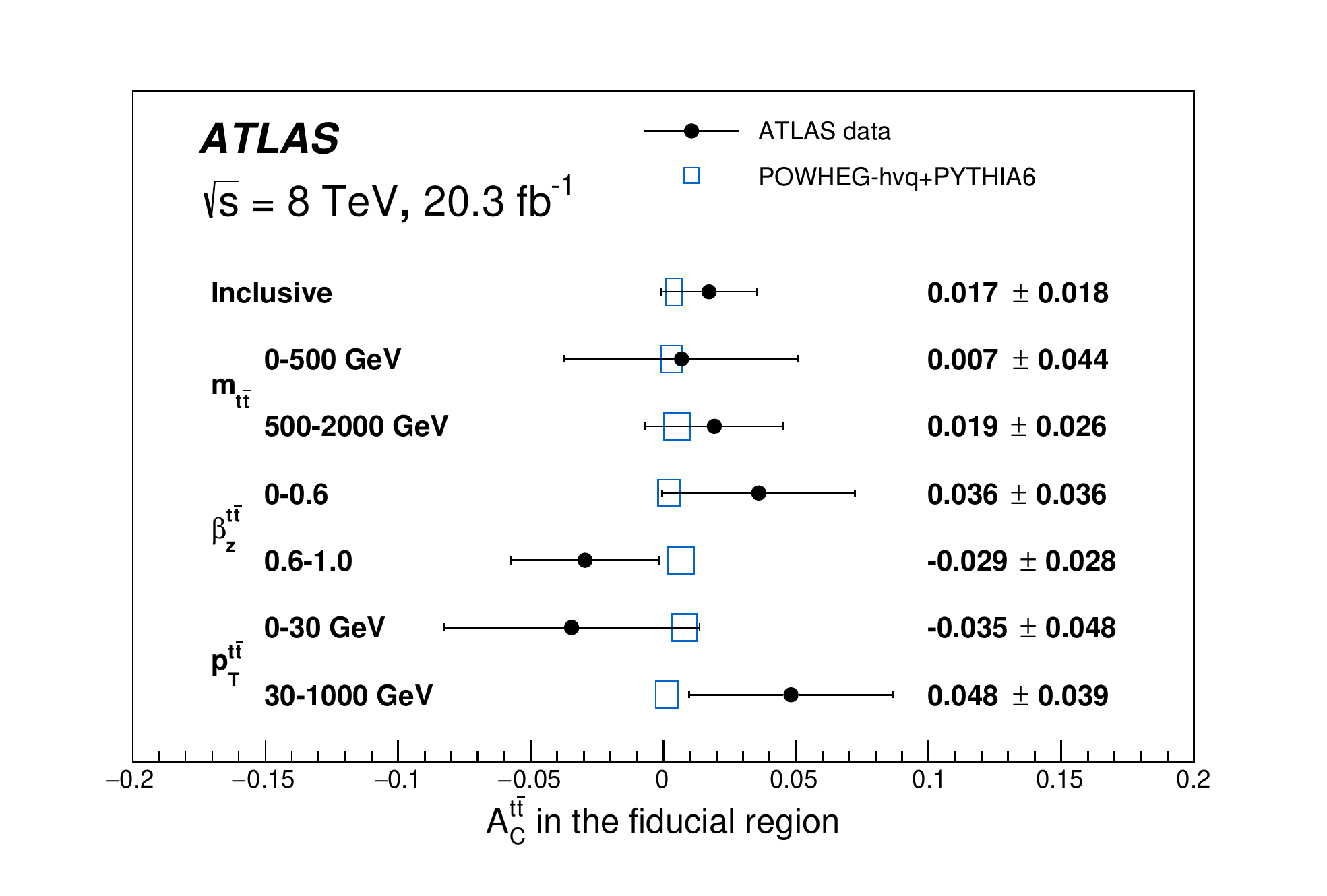}
\includegraphics[width=0.70\columnwidth]{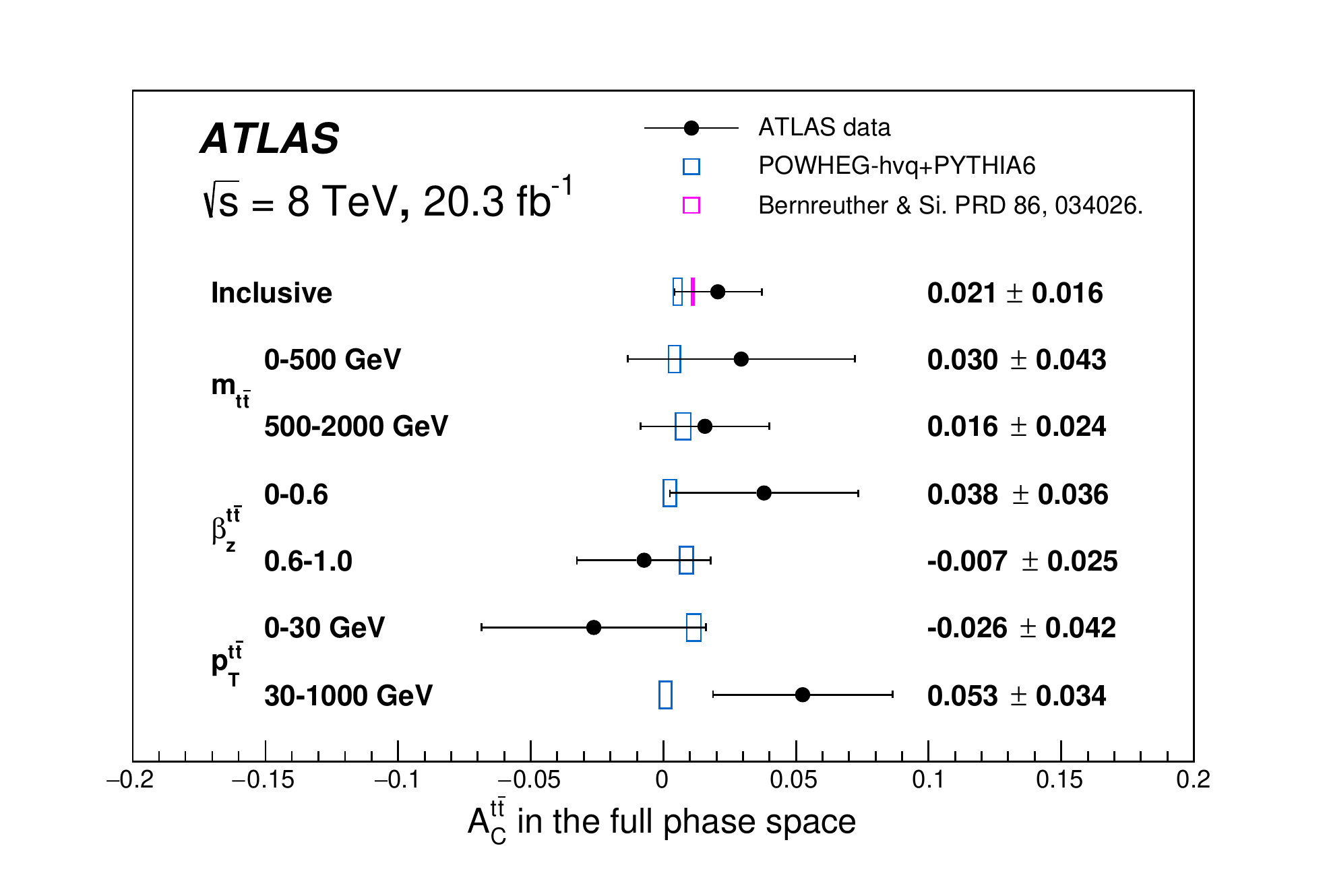}

\caption{Summary of all the measurements in this paper for the $\ttbar$
asymmetry in the fiducial volume (top) and  full phase space
(bottom). The predictions shown in blue are obtained using \powheg
+ \pythia at NLO where the uncertainties are statistical, and the
corresponding theoretical uncertainties are
small compared to the experimental precision. The inclusive
measurement in the full phase space is compared to a NLO + EW
prediction ~\cite{Bernreuther:2012sx}.}
\label{fig:sumary_plot_ttbar}
\end{center}
\end{figure}

\begin{figure}[h]
\begin{center}
\includegraphics[width=0.45\columnwidth]{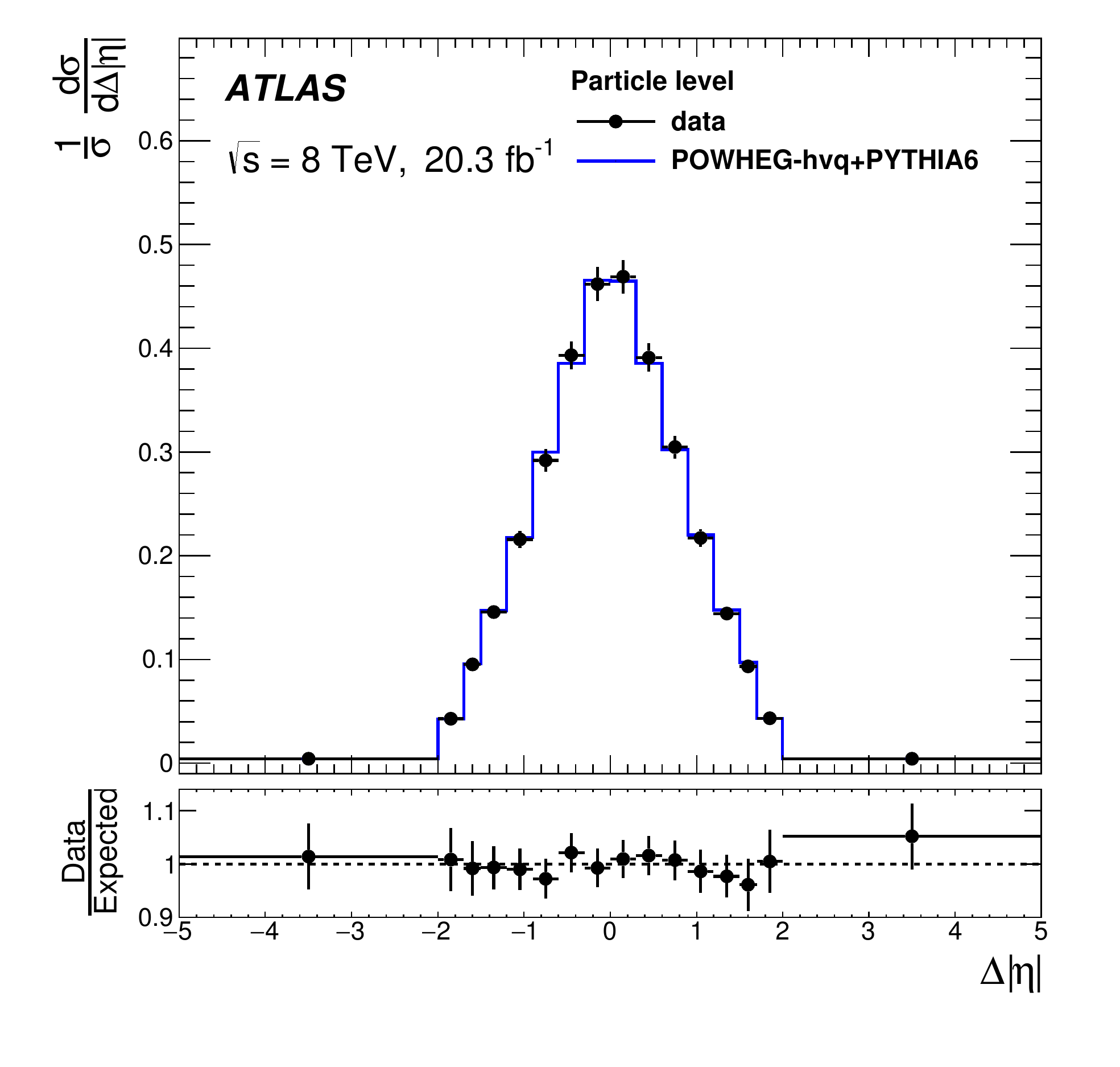}
\includegraphics[width=0.45\columnwidth]{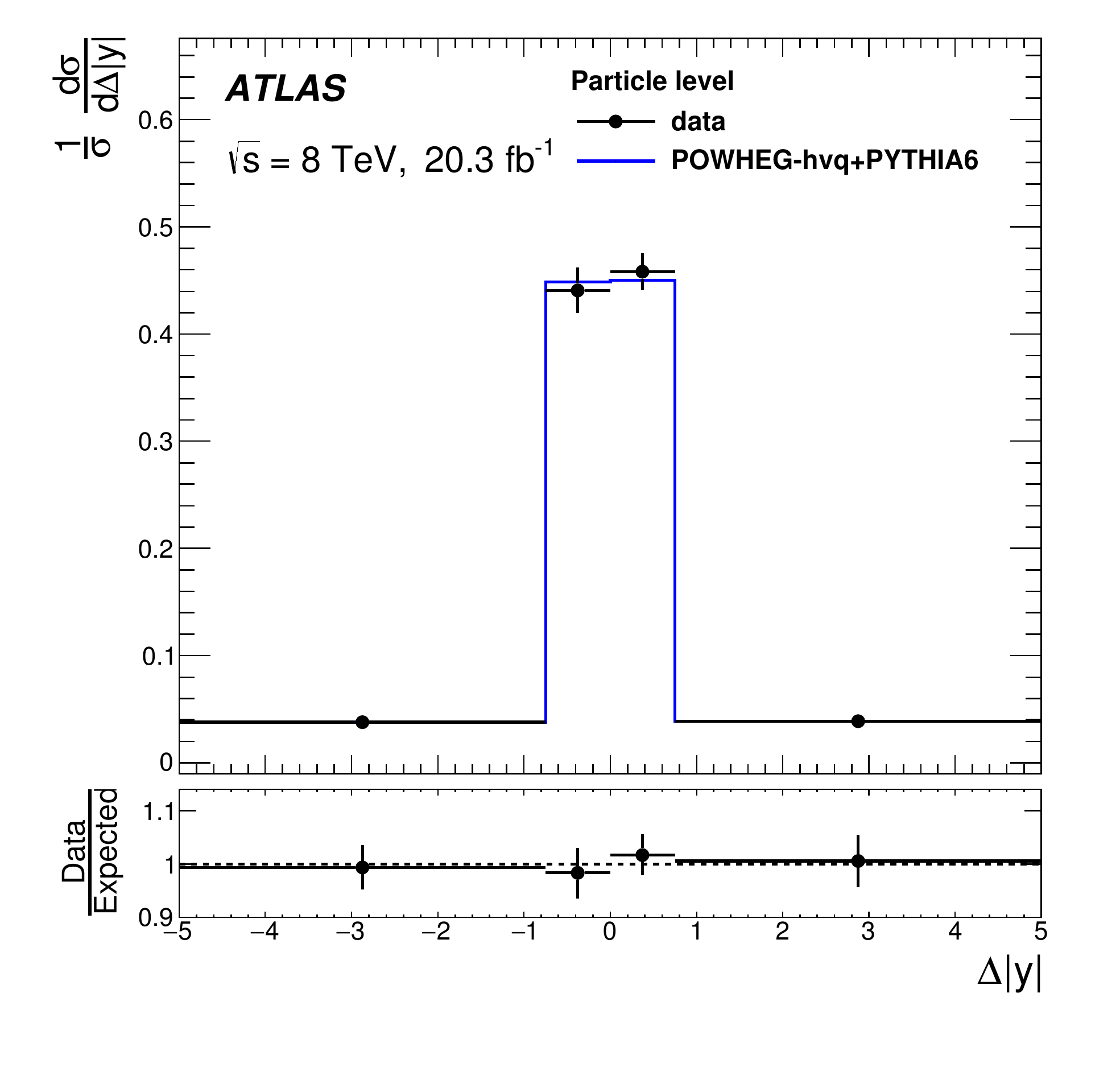}
\caption{Data distribution after the unfolding procedure compared with the \powheg + \pythia prediction at
NLO for the inclusive $\Delta|\eta|$ (left) and $\Delta|y|$ (right) observables in the fiducial volume. The data/expected ratio is also shown.}
\label{fig:unfolded_plots}
\end{center}
\end{figure}

Figures~\ref{fig:sumary_plot_lepton} and~\ref{fig:sumary_plot_ttbar}
show the inclusive and differential results for the leptonic and
$t\bar{t}$ charge asymmetry in the fiducial region and in the full
phase space.  All the results are compatible with the Standard Model
predictions~\cite{Bernreuther:2012sx,Alioli:2010xd,Frixione:2007vw,Nason:2004rx}.
Figure~\ref{fig:unfolded_plots} shows the unfolded
distributions of the $\Delta|\eta|$  and $\Delta|y|$  observables for the
inclusive measurement in the fiducial volume.
 The distributions are compared with Monte
Carlo predictions at NLO provided by \powheg. The measured inclusive
values in the full phase space are $\Acll =  \Acllcomb$ and $\Ac
= \Accomb$.
They are in agreement with
the Standard Model predictions \mbox{$\Acll = \Aclltheory$}
and \mbox{$\Actt = \Actttheory$}~\cite{Bernreuther:2012sx}.
The measurements are consistent with other LHC asymmetry
measurements at 8 \TeV~\cite{cms8tevljets,
atlas8tevljets,Khachatryan:2015mna}.

The statistical uncertainty is in
most cases the dominant contribution to the total uncertainty.
The dominant systematic uncertainties across all the measurements are the signal modeling and the kinematic reconstruction uncertainty.
The signal modeling uncertainties are reduced in most of the cases by performing the measurements in the fiducial region,
since the extrapolation from detector acceptance to the full phase space is avoided.
The statistical uncertainty is slightly larger in the fiducial region than in the full phase space;
this is expected because some reconstructed events fail the fiducial requirements in the fiducial analysis.

\begin{figure}[h]
\begin{center}
\includegraphics[width=0.45\columnwidth]{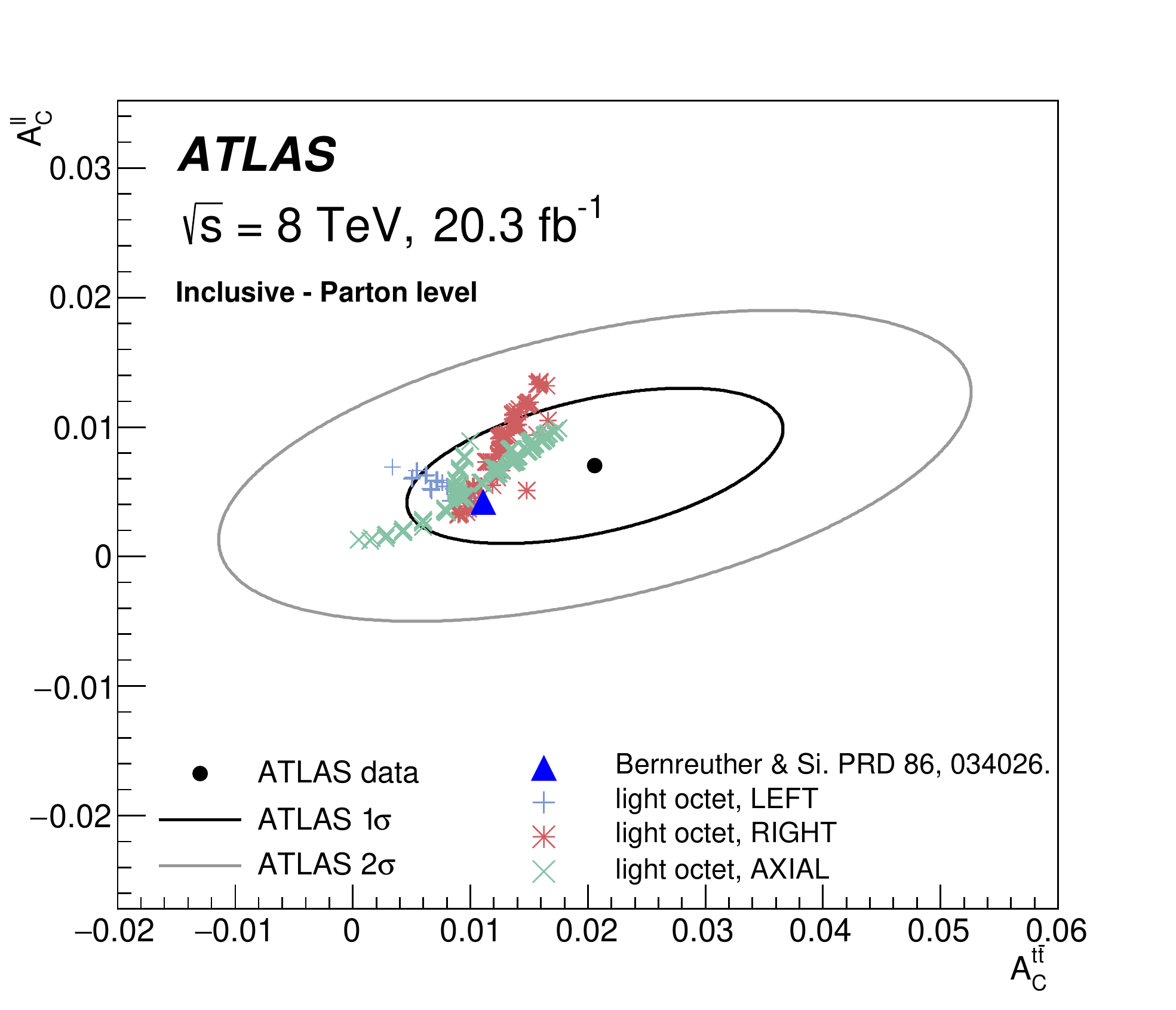}
\includegraphics[width=0.45\columnwidth]{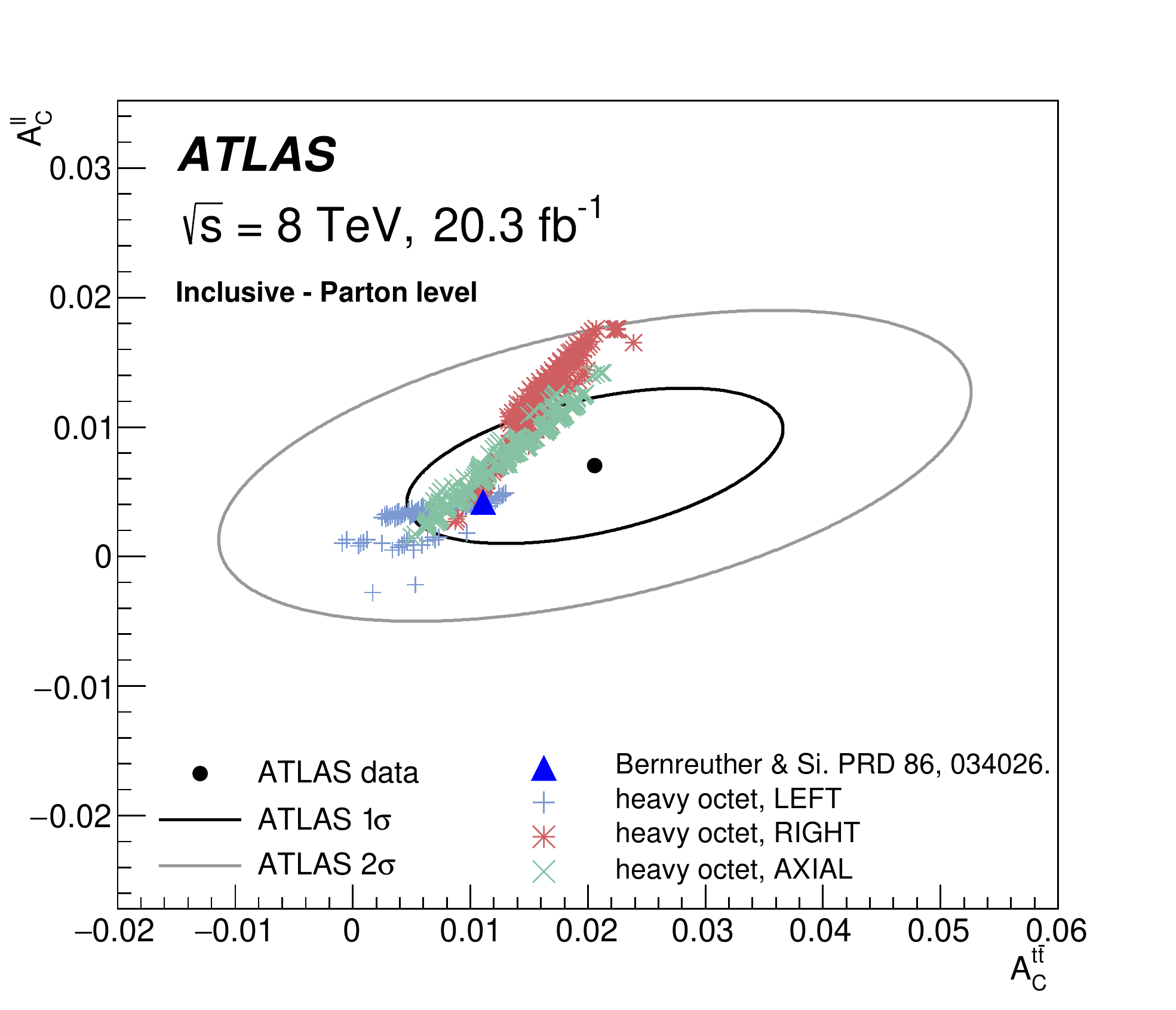}
\caption{Comparison of the inclusive $A_{\rm{C}}^{\ell \ell}$ and
$A_{\rm{C}}^{t\bar{t}}$ measurement values  in the full phase space to
the SM NLO QCD+EW prediction~\cite{Bernreuther:2012sx} and to two benchmark
BSM models~\cite{Aguilar-Saavedra:2014nja}, one with a light octet with mass below the $t\bar{t}$ production threshold (left) and one with
a heavy octet with mass beyond the reach of the LHC (right), for various couplings as described in
the legend. Ellipses corresponding to $1\sigma$ and $2\sigma$ combined statistical and
systematic uncertainties of the measurement, including the correlation between $A_{\rm{C}}^{\ell \ell}$ and $A_{\rm{C}}^{t\bar{t}}$, are also shown. }
\label{fig:2d_map_dy}
\end{center}
\end{figure}

Figure~\ref{fig:2d_map_dy} compares the values of
$A_{\rm{C}}^{\ell \ell}$ and $A_{\rm{C}}^{t\bar{t}}$ from the
inclusive measurements in the full phase space
to the SM  predictions and two BSM models~\cite{Aguilar-Saavedra:2014nja}
 compatible with the Tevatron results.
 Two BSM models with a new color-octet particle that is exchanged in the
 $s$-channel are considered. In the model with the light octet, the
 new particle's mass ($m = 250$ \GeV) is below the $t\bar{t}$
production threshold and its width is assumed to be $\Gamma = 0.2m$. The model with the heavy octet uses an octet
 mass beyond current limits from direct searches at the LHC. The
 corrections to $t\bar{t}$ production are independent of
the mass but instead depend on the ratio of coupling to mass, which is
assumed to be $1 \, \rm{\TeV}^{-1}$. The new particles in both BSM models would
not be visible as resonances in the $m_{t\bar{t}}$ spectrum at the
Tevatron or at the LHC.
In the figures, model predictions for different left-handed, right-handed, and axial coupling constants
to top quarks are shown. The ellipses correspond to
the $1 \sigma$ and $2 \sigma$ total uncertainty in the measurements. The correlation between these
two measurements is taken into account. The statistical and detector
systematic uncertainty correlation between $A_{\rm{C}}^{\ell \ell}$ and $A_{\rm{C}}^{t\bar{t}}$
is found to be $30 \%$. The modeling systematic uncertainties are assumed
to be $100\%$ correlated.  The resulting correlation between
$A_{\rm{C}}^{\ell \ell}$ and $A_{\rm{C}}^{t\bar{t}}$  is about $48\%$.
The measurements are compatible with the SM and do not exclude the two sets of BSM models considered.

\FloatBarrier

\section{Conclusion}
\label{sec:conclusions}
Measurements of the leptonic and \ttbar{} charge asymmetry in the
dilepton channel, characterized
by two high-$p_{\rm T}$ leptons (electrons or muons), are presented. The measurements, corrected for
detector resolution and acceptance effects, are performed
using data corresponding to an integrated luminosity of \lumi of \pp
collisions at $\sqrt{s} = 8$~\tev\ collected by the \atlas
detector at the LHC. The inclusive asymmetries are measured in the full
phase space to be:
\begin{eqnarray*}
  \Acll & = & \Acllcomb~\textrm{and}
\end{eqnarray*}
\begin{eqnarray*}
   \Ac & = & \Accomb.
\end{eqnarray*}
They are in agreement with the Standard Model predictions $\Acll
= \Aclltheory$ and $\Ac = \Actttheory$.
Differential measurements of the asymmetries as a function
of the invariant mass, transverse momentum, and longitudinal boost of
the $t\bar{t}$  system are also performed and they are found to be in
agreement with the SM predictions, although they have relatively large
 uncertainties.
All measurements are also performed in a fiducial region at particle
level where the modeling uncertainties are reduced.
For all measurements, the statistical uncertainty is the
dominant contribution to the total uncertainty. The unfolded distributions of
lepton \deta and \ttbar{} \dy are provided. Good agreement between
the corrected distributions and the predictions of \powheg + \pythia is
observed.


\section*{Acknowledgments}
We thank CERN for the very successful operation of the LHC, as well as the
support staff from our institutions without whom ATLAS could not be
operated efficiently.

We acknowledge the support of ANPCyT, Argentina; YerPhI, Armenia; ARC, Australia; BMWFW and FWF, Austria; ANAS, Azerbaijan; SSTC, Belarus; CNPq and FAPESP, Brazil; NSERC, NRC and CFI, Canada; CERN; CONICYT, Chile; CAS, MOST and NSFC, China; COLCIENCIAS, Colombia; MSMT CR, MPO CR and VSC CR, Czech Republic; DNRF and DNSRC, Denmark; IN2P3-CNRS, CEA-DSM/IRFU, France; GNSF, Georgia; BMBF, HGF, and MPG, Germany; GSRT, Greece; RGC, Hong Kong SAR, China; ISF, I-CORE and Benoziyo Center, Israel; INFN, Italy; MEXT and JSPS, Japan; CNRST, Morocco; FOM and NWO, Netherlands; RCN, Norway; MNiSW and NCN, Poland; FCT, Portugal; MNE/IFA, Romania; MES of Russia and NRC KI, Russian Federation; JINR; MESTD, Serbia; MSSR, Slovakia; ARRS and MIZ\v{S}, Slovenia; DST/NRF, South Africa; MINECO, Spain; SRC and Wallenberg Foundation, Sweden; SERI, SNSF and Cantons of Bern and Geneva, Switzerland; MOST, Taiwan; TAEK, Turkey; STFC, United Kingdom; DOE and NSF, United States of America. In addition, individual groups and members have received support from BCKDF, the Canada Council, CANARIE, CRC, Compute Canada, FQRNT, and the Ontario Innovation Trust, Canada; EPLANET, ERC, FP7, Horizon 2020 and Marie Sk{\l}odowska-Curie Actions, European Union; Investissements d'Avenir Labex and Idex, ANR, R{\'e}gion Auvergne and Fondation Partager le Savoir, France; DFG and AvH Foundation, Germany; Herakleitos, Thales and Aristeia programmes co-financed by EU-ESF and the Greek NSRF; BSF, GIF and Minerva, Israel; BRF, Norway; Generalitat de Catalunya, Generalitat Valenciana, Spain; the Royal Society and Leverhulme Trust, United Kingdom.

The crucial computing support from all WLCG partners is acknowledged gratefully, in particular from CERN, the ATLAS Tier-1 facilities at TRIUMF (Canada), NDGF (Denmark, Norway, Sweden), CC-IN2P3 (France), KIT/GridKA (Germany), INFN-CNAF (Italy), NL-T1 (Netherlands), PIC (Spain), ASGC (Taiwan), RAL (UK) and BNL (USA), the Tier-2 facilities worldwide and large non-WLCG resource providers. Major contributors of computing resources are listed in Ref.~\cite{ATL-GEN-PUB-2016-002}.


\printbibliography

 \newpage 
\begin{flushleft}
{\Large The ATLAS Collaboration}

\bigskip

G.~Aad$^\textrm{\scriptsize 87}$,
B.~Abbott$^\textrm{\scriptsize 114}$,
J.~Abdallah$^\textrm{\scriptsize 65}$,
O.~Abdinov$^\textrm{\scriptsize 12}$,
B.~Abeloos$^\textrm{\scriptsize 118}$,
R.~Aben$^\textrm{\scriptsize 108}$,
M.~Abolins$^\textrm{\scriptsize 92}$,
O.S.~AbouZeid$^\textrm{\scriptsize 138}$,
N.L.~Abraham$^\textrm{\scriptsize 150}$,
H.~Abramowicz$^\textrm{\scriptsize 154}$
% ...
\bigskip
\\
% ...
$^{12}$ Institute of Physics, Azerbaijan Academy of Sciences, Baku, Azerbaijan\\
% ...
$^{65}$ University of Iowa, Iowa City IA, United States of America\\
% ...
$^{87}$ CPPM, Aix-Marseille Universit{\'e} and CNRS/IN2P3, Marseille, France\\
% ...
$^{92}$ Department of Physics and Astronomy, Michigan State University, East Lansing MI, United States of America\\
% ...
$^{108}$ Nikhef National Institute for Subatomic Physics and University of Amsterdam, Amsterdam, Netherlands\\
% ...
$^{114}$ Homer L. Dodge Department of Physics and Astronomy, University of Oklahoma, Norman OK, United States of America\\
% ...
$^{118}$ LAL, Univ. Paris-Sud, CNRS/IN2P3, Universit{\'e} Paris-Saclay, Orsay, France\\
% ...
$^{138}$ Santa Cruz Institute for Particle Physics, University of California Santa Cruz, Santa Cruz CA, United States of America\\
% ...
$^{150}$ Department of Physics and Astronomy, University of Sussex, Brighton, United Kingdom\\
% ...
$^{154}$ Raymond and Beverly Sackler School of Physics and Astronomy, Tel Aviv University, Tel Aviv, Israel\\
% ...
\end{flushleft}



\end{document}